\documentclass[preprint,pra,nobalancelastpage,aps,raggedbottom, superscriptaddress]{revtex4-2}
\usepackage{times}

\usepackage{amsmath,amsfonts,amssymb,graphicx,hyperref,epstopdf,xcolor,tikz,scalerel,natbib,array}
\usepackage{mathptmx,textcomp,braket}
\usepackage[T1]{fontenc}
\usepackage[utf8]{inputenc}
\usepackage{multirow}
\usepackage{array}

\begin{document}

\title{Generation of Circularly-Polarised  High-Harmonics with Identical  Helicity in Two-Dimensional Materials}

\author{Navdeep Rana}
\affiliation{%
Department of Physics, Indian Institute of Technology Bombay,
           Powai, Mumbai 400076, India }

\author{M. S. Mrudul}
\affiliation{%
Department of Physics and Astronomy, Uppsala University, P.O. Box 516, SE 75120, Uppsala, Sweden }
                     
\author{Gopal Dixit}
\email[]{gdixit@phy.iitb.ac.in}
\affiliation{%
Department of Physics, Indian Institute of Technology Bombay, Powai, Mumbai 400076, India }

\date{\today}


\begin{abstract}
Generation of circularly-polarized high-harmonics with the same helicity to all orders is indispensable for 
chiral-sensitive spectroscopy with attosecond temporal resolution. Solid-state samples have added a valuable 
asset in controlling the polarization of emitted harmonics. 
However, maintaining the identical helicity of the emitted harmonics to all orders is a daunting task.  
In this work, we  demonstrate a robust recipe for efficient 
generation of circularly-polarized harmonics with the same helicity. For this purpose, a nontrivial tailored 
driving field, consisting of two co-rotating laser pulses with frequencies $\omega$ and $2\omega$, is utilized 
to generate harmonics from graphene. The Lissajous figure of the total driving pulse exhibits an absence of the 
rotational symmetry, which imposes no constraint on the helicity of the emitted harmonics. Our approach 
to generating circularly-polarized harmonics with the same helicity is robust against various perturbations in the 
setup, such as variation in the subcycle phase difference  
or the intensity ratio of the $\omega$ and $2\omega$  pulses, as 
rotational symmetry of the total 
driving pulse remains absent. Our approach is expected to be equally applicable to 
other two-dimensional materials, among others,  transition-metal dichalcogenides and hexagonal boron nitride as our approach is  based on  absence of the rotational symmetry of the driving pulse. 
Our work paves the way for establishing 
compact solid-state chiral-XUV sources, opening a new realm for chiral light-matter interaction on its intrinsic timescale. 
\end{abstract}

\maketitle 

\newpage 

\section{Introduction} 
In the last few decades, high-harmonic generation (HHG) has become an essential ingredient in attosecond science. 
HHG is a strong-field-driven highly  nonlinear frequency up-conversion process.  
HHG provides a route not only to generate attosecond pulses in the extreme ultraviolet  (XUV)  
energy regime~\cite{midorikawa2022progress} 
but also to interrogate electron dynamics on its natural 
timescale~\cite{krausz2009attosecond, bucksbaum2007, corkum2007, dixit2012}.  
Owing to the recollision mechanism of HHG in gases~\cite{corkum1993plasma}, 
the harmonic yield drops markedly  as the polarization 
of the driving laser pulse changes from linear to circular. 
To circumvent this problem, a combination of two counter-rotating circularly-polarized laser pulses  
with frequencies $\omega$ and $2\omega$ has been employed to generate  circularly-polarized harmonics. 
The resultant harmonic spectrum  displays doublets of circularly-polarized harmonics where 
$(3n+1)$ and $(3n+2)$ harmonics exhibit the polarization of
$\omega$- and $2\omega$-fields, respectively.  However,  $3n$ harmonics do not conserve parity and 
thus they are  forbidden~\cite{neufeld2019floquet, fleischer2014spin}.
Different variants of the bichromatic counter-rotating $\omega-2\omega$ schemes, such as 
varying  the ratio of the ellipticities and/or of the intensities of the two driving pulses~\cite{neufeld2018optical, dorney2017helicity}, controlling the subcycle phase between the two pulses~\cite{frolov2018control}, 
introduction of an additional seed pulse with different  polarization~\cite{dixit2018control, rajpoot2021polarization},  
adding  plasmonic field enhancement~\cite{ansari2021controlling}, 
and noncollinear geometry of 
the $\omega-2\omega$ scheme~\cite{huang2018polarization, hickstein2015non} 
are employed to control the polarization of the emitted harmonics. 
It is important to emphasis that the recognition of chiral molecules~\cite{ferre2015table, cireasa2015probing}, 
circular dichroism in magnetic materials~\cite{kfir2015generation}, 
spin and magnetisation dynamics in solids at their natural timescales~\cite{boeglin2010distinguishing, radu2011transient},  to name but a few, are 
examples of chiral-sensitive  light-matter interaction  phenomena where 
the desired control over the polarization of the emitted harmonics and resultant 
attosecond XUV pulses is quintessential.

The extension of HHG from gases to solids offers an attractive 
option for compact tabletop sources of coherent XUV attosecond pulses. 
Because of the periodic nature of solids with higher electron density in comparison with gases, 
HHG in solids  seems a better option for higher yield of harmonics~\cite{goulielmakis2022high, ghimire2019}. 
Moreover, solid HHG has added a new dimension in attosecond spectroscopy to shed new light on various equilibrium and  nonequilibrium 
aspects of solids~\cite{bharti2022high, pattanayak2022role, zaks2012experimental, luu2015extreme, schubert2014sub,  hohenleutner2015real, pattanayak2020influence, vampa2015linking, vampa2015all, langer2018lightwave, luu2018measurement, banks2017dynamical, pattanayak2019direct, imai2020high, borsch2020super, rana2022high}. 

In comparison with HHG from gases,  HHG from solids exhibits different sensitivity  toward 
circularly-polarized laser pulses~\cite{ghimire2011observation, tancogne2017ellipticity}.  
It has been shown that 
a high degree of control over the polarization of emitted 
harmonics can be achieved  by the 
dynamic control of crystal symmetries and intertwined 
interband and intraband electronic dynamics~\cite{klemke2019polarization}. 
Saito \textit{et al.} have demonstrated the generation of 
harmonics with alternate helicity using 
a circularly-polarized laser pulse,  owing to the selection rules derived from the symmetry~\cite{saito2017observation}. 
Recently, bicircular counter-rotating $\omega-2\omega$ laser pulses have been employed to generate 
circularly-polarized harmonics in solids, which has enabled  one to preform symmetry-resolved chiral spectroscopy~\cite{heinrich2021chiral} and to probe valley-selective excitations in two-dimensional materials ~\cite{he2022dynamical, mrudul2021light, mrudul2021controlling}.  
In all cases, the harmonic spectra of solids, similar to gases, are composed of pairs of circularly-polarized harmonics with opposite helicity  owing to the threefold rotational symmetry of the bicircular counter-rotating pulses. 
Thus, the generation of  circularly-polarized harmonics with the same helicity to all orders is missing -- a major impediment in the realisation of a compact all-solid-state XUV source for chiral spectroscopy. 
The main focus of the present work is to address this major obstacle.

Our work provides a robust recipe to generate  
circularly-polarized harmonics with the same helicity to all orders. 
For that purpose, a nontrivial tailored driving field without any rotational symmetry  is designed for HHG. 
The total electric field of the co-rotating bicircular $\omega-2\omega$ pulses does not exhibit any  symmetry, which is in contrast to the counter-rotating  $\omega-2\omega$ pulses with threefold symmetry. 
Thus,  it is expected that the absence of any symmetry of the total driving field  will impose 
no symmetry constraint on  the helicity of the emitted harmonics. 
To test our idea,  monolayer graphene is exposed to a combination of  bicircular co-rotating $\omega-2\omega$ laser pulses. 
It is worth mentioning that HHG from graphene and the underlying mechanism have been
extensively explored in recent years~\cite{hafez2018extremely, avetissian2018impact, chizhova2017high, al2014high,  zurron2018theory, zurron2019optical, taucer2017nonperturbative, chen2019circularly,   sato2021high, boyero2022non}. 
Moreover,  the anomalous ellipticity dependence of HHG from graphene has been
explored experimentally~\cite{yoshikawa2017high, taucer2017nonperturbative} as well as discussed theoretically~\cite{yoshikawa2017high, taucer2017nonperturbative, zhang2021orientation, liu2018driving, dong2021ellipticity, zurron2019optical}.  

\section{Computational Details}
Semiconductor Bloch equations in the Houston basis are solved to simulate the interaction of laser with graphene. 
The  nearest-neighbor tight-binding approach is used to represent graphene as discussed in our previous work~\cite{mrudul2021high}. 
High-harmonic spectrum is  simulated by performing the 
Fourier transform  ($\mathcal{FT}$) of the time derivative of the total current as	
\begin{equation}
\mathcal{I}(\omega) = \left|\mathcal{FT}\left(\frac{d}{dt} \left[\int_{BZ} \textbf{J}(\mathbf{k}, t)~d{\textbf{k}} 
\right] \right) \right|^2.
\end{equation}
Here, $\textbf{J}(\mathbf{k}, t)$ is the current at any $\textbf{k}$-point 
in the Brillouin zone as a function of time. 

\section{Results and Discussion}
\begin{figure}
\includegraphics[width= 0.8 \linewidth]{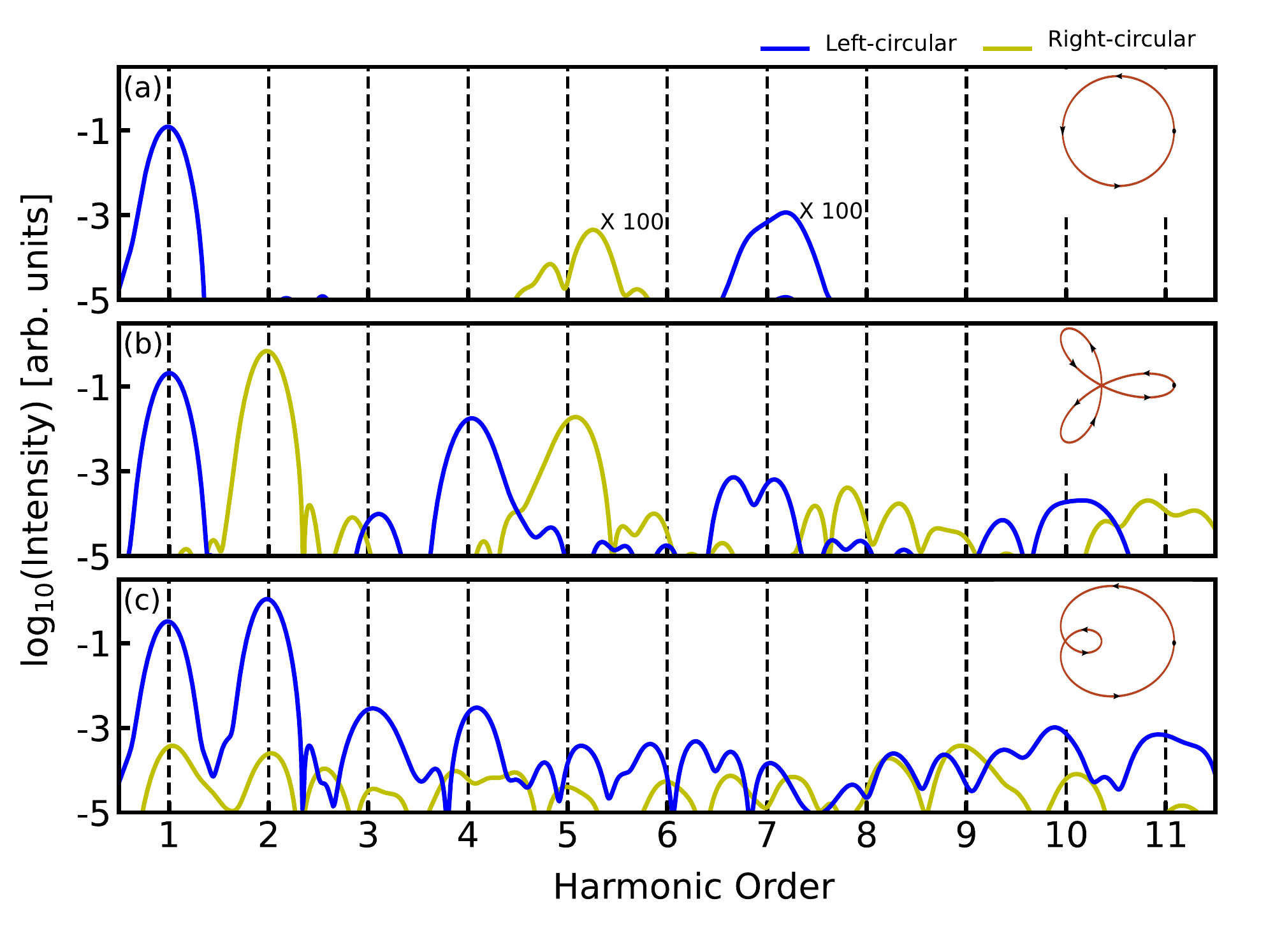}
\caption{Polarization-resolved high-harmonic spectrum of graphene driven by (a) single-colour 
circularly-polarized laser,   
(b) bicircular counter-rotating $\omega$-2$\omega$, and  (c)  bicircular co-rotating $\omega$-2$\omega$  laser pulses. 
In all cases, the fundamental $\omega$ pulse is left-handed polarized. 
Lissajous figures of the total driving fields are shown in the respective insets. 
In all cases, the  peak intensity of  the $\omega$ pulse is  0.3 TW/cm$^{2}$  and the wavelength is 2 $\mu$m. 
The subcycle phase difference between the two pulses  $\phi$ is zero, 
and  the relative strength of the electric fields between $\omega$ and 2$\omega$ pulses $\mathsf{R}$ 
is 2 in (b) and (c).} \label{Spectra}
\end{figure}

Before we discuss results obtained for bicircular co-rotating $\omega-2\omega$ laser pulses, let us revisit
the results for single-colour and  bicircular counter-rotating $\omega-2\omega$ laser pulses.
The polarization-resolved high-harmonic spectrum of graphene driven by a single-colour 
left-handed circularly-polarized pulse is shown in Fig.~$\ref{Spectra}$(a). 
Owing to the six-fold rotational symmetry of graphene and conservation of spin angular momentum of the driving laser pulse, selection rules indicate the generation of 
(6$n$ $\pm$1)-order harmonics with $n = 0, 1, 2, 3, \cdots$. 
Moreover, the helicity of the 6$n$+1 (6$n$-1) harmonics is the 
same (opposite)  helicity as the driving pulse~\cite{alon1998selection}.  
Our results in Fig.~$\ref{Spectra}$(a) are in prefect agreement  with the selection rules and with an 
earlier report~\cite{chen2019circularly}. 
Moreover,  the low harmonic yield  of the fifth and seventh harmonics with opposite helicity has  
been observed in  an experiment~\cite{yoshikawa2017high}. 
It has been anticipated that the yield of harmonics can be improved by 
increasing the intensity of the driving field~\cite{chen2019circularly}. 

The total vector potential corresponding to $\omega-2\omega$ circularly-polarized laser pulses is given as
\begin{equation}\label{eq1}
\textbf{A}(t) = \frac{A_{0}f(t)}{\sqrt{2}}\left( \left[ \cos(\omega t + \phi) + \frac{\textsf{R}}{2}\cos(2 \omega t)\right]\hat{\textbf{e}}_{x}+\left[ \sin(\omega t + \phi)\pm \frac{\textsf{R}}{2}\sin(2 \omega t)\right]\hat{\textbf{e}}_{y}\right).
\end{equation}
Here, $A_{0}$ is the amplitude of the vector potential, $f(t)$ is the temporal envelope of the driving field, 
$\phi$ is the subcycle phase difference 
between  $\omega$ and 2$\omega$ pulses, 
and $\mathsf{R}$ is the ratio of electric field strength of the two pulses.  +(-) represents the co-rotating (counter-rotating) laser pulse configuration.  
In this work, fundamental  $\omega$ pulse with a peak intensity
of 0.3 TW/cm$^{2}$ and a wavelength of 2 $\mu$m is used for HHG from graphene. 
The fundamental driving pulse  has eight cycles with sin-square envelope.
Similar laser parameters have been employed to study coherent electron dynamics in graphene~\cite{higuchi2017light, heide2018coherent}.

Figure $\ref{Spectra}$(b) presents the polarization-resolved harmonic spectrum of graphene driven by 
 bicircular  counter-rotating $\omega$-2$\omega$ laser pulses. 
The resultant vector potential exhibits  
a  trefoil symmetry as shown in the inset. 
Following the threefold symmetry and conservation of  spin angular momentum, 
$n\omega = p\omega + 2q\omega$  with  $p = q \pm 1$  harmonics are allowed, whereas  
$3n$ harmonics are symmetry forbidden~\cite{fleischer2014spin}.
 $p (q)$  is  the number of photons of the $\omega (2\omega)$ pulse. 
Moreover,  the allowed  $(3n + 1)$- and $(3n + 2)$-order harmonics follow the helicity of the $\omega$ and 
$2\omega$  pulses, respectively~\cite{kfir2015generation, heinrich2021chiral, he2022dynamical}. 
From the figure, our results are consistent with the selection rules and earlier report~\cite{mrudul2021light}.   
Note that the polarization of one of the driving pulses individually impacts 
the helicity of the emitted  harmonics significantly. 
 
Analysis of results in Figs~$\ref{Spectra}$(a) and $\ref{Spectra}$(b) establishes  that  
the generation of  harmonics with circular polarization, using either a single colour circular pulse 
or  bicircular  counter-rotating $\omega$-2$\omega$ pulses, is possible. 
In both cases, the adjacent harmonics are of opposite helicity, which 
has a serious consequence for the helicity of the resultant   attosecond pulses. 
If the intensities of adjacent harmonics with opposite helicity are the same then the generated attosecond pulse
exhibits linearly polarization  with a rotating axis of polarization. 
In such circumstances, one needs to induce an imbalance in the intensity of the  adjacent harmonics to have a control over the helicity of the resultant 
attosecond pulses~\cite{dorney2017helicity, frolov2018control, dixit2018control}. 
 
The total harmonic spectrum in polarization-resolved fashion for the bicircular  
co-rotating $\omega$-2$\omega$ laser pulses is presented in Fig.~\ref{Spectra}(c). 
As evident from the Lissajous figure in the inset, the total field does not exhibit  any rotational symmetry. Thus, 
the conservation of  spin angular momentum does not impose any constraint in the helicity of the generated harmonics and none of the harmonics are symmetry forbidden. As a result, all-order circularly-polarized harmonics  are generated with the same helicity as that of the driving pulse.  
Analysis of the relative yield of the spectrum revels that the left-circularly-polarized harmonics are  
the leading harmonics.

\begin{figure}
\includegraphics[width=\linewidth]{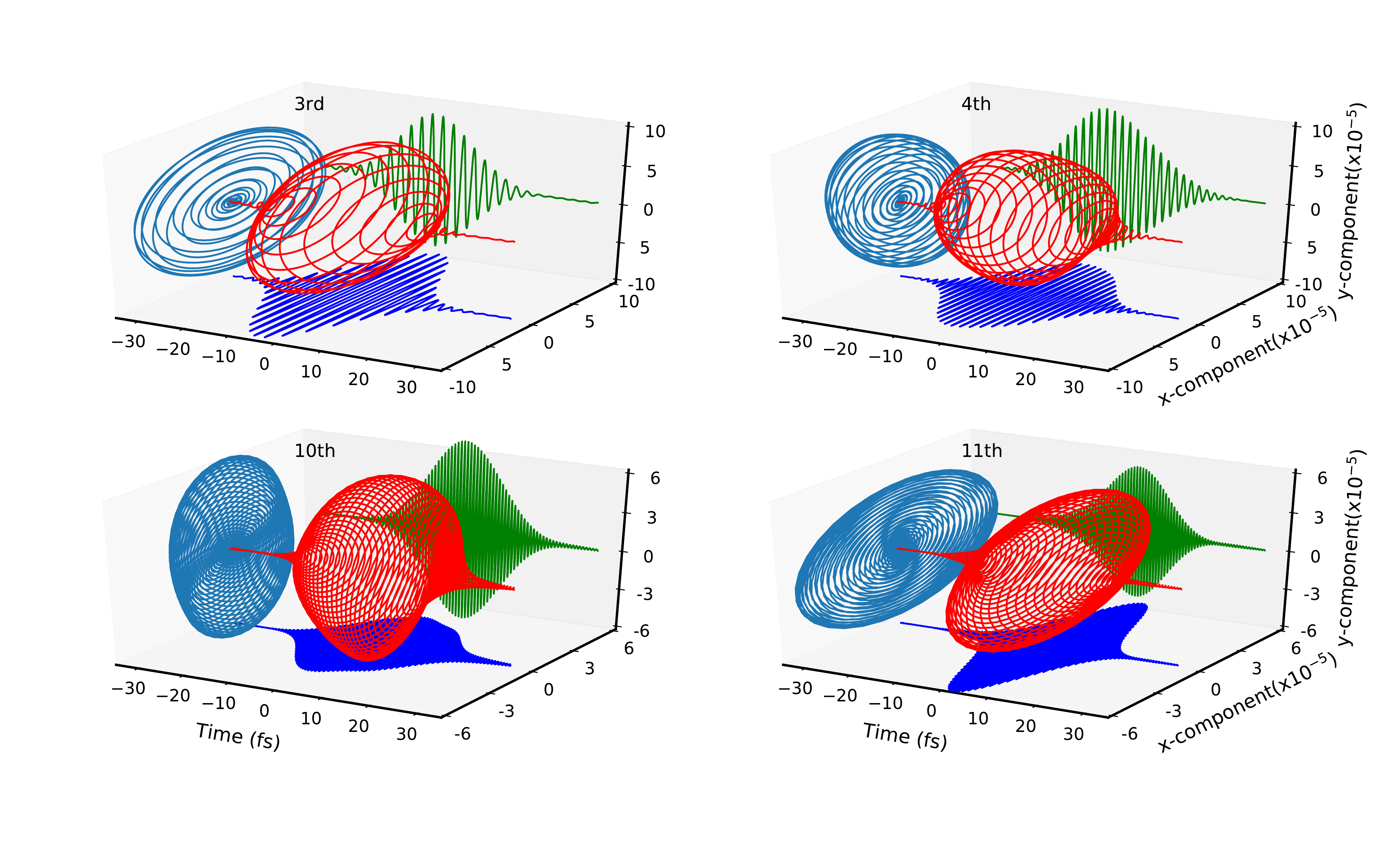}
\caption{Time profiles of the emitted harmonics, shown in  Fig.~\ref{Spectra}(c), generated by   bicircular co-rotating $\omega$-2$\omega$  laser pulses (red color). 
The $x$ and $y$ components of the harmonics are shown in blue and green colours, respectively. 
The Lissajous figures of the emitted harmonics are shown in cyan colour. All harmonics are left-handed circularly polarized. The ellipticities of the 3$^{\textrm{rd}}$, 4$^{\textrm{th}}$, 10$^{\textrm{th}}$ and 11$^{\textrm{th}}$ harmonics are 0.8, 0.8, 0.8 and 0.9, respectively.} 
\label{harmonics}
\end{figure}

After demonstrating the capability of the co-rotating $\omega$-2$\omega$ scheme to generate 
circularly-polarized harmonics with the same helicity, let us analysis the temporal evolution 
of the emitted harmonics in the time domain. 
For this purpose, Fig.~\ref{harmonics} presents the time profiles of a few selected harmonics corresponding to Fig.~\ref{Spectra}(c).
All the harmonics are left-handed circularly polarized 
as evident from the Lissajous figure and the  $x$ and $y$ components of the harmonics in the time domain. 

\begin{figure}
\includegraphics[width= 0.9\linewidth]{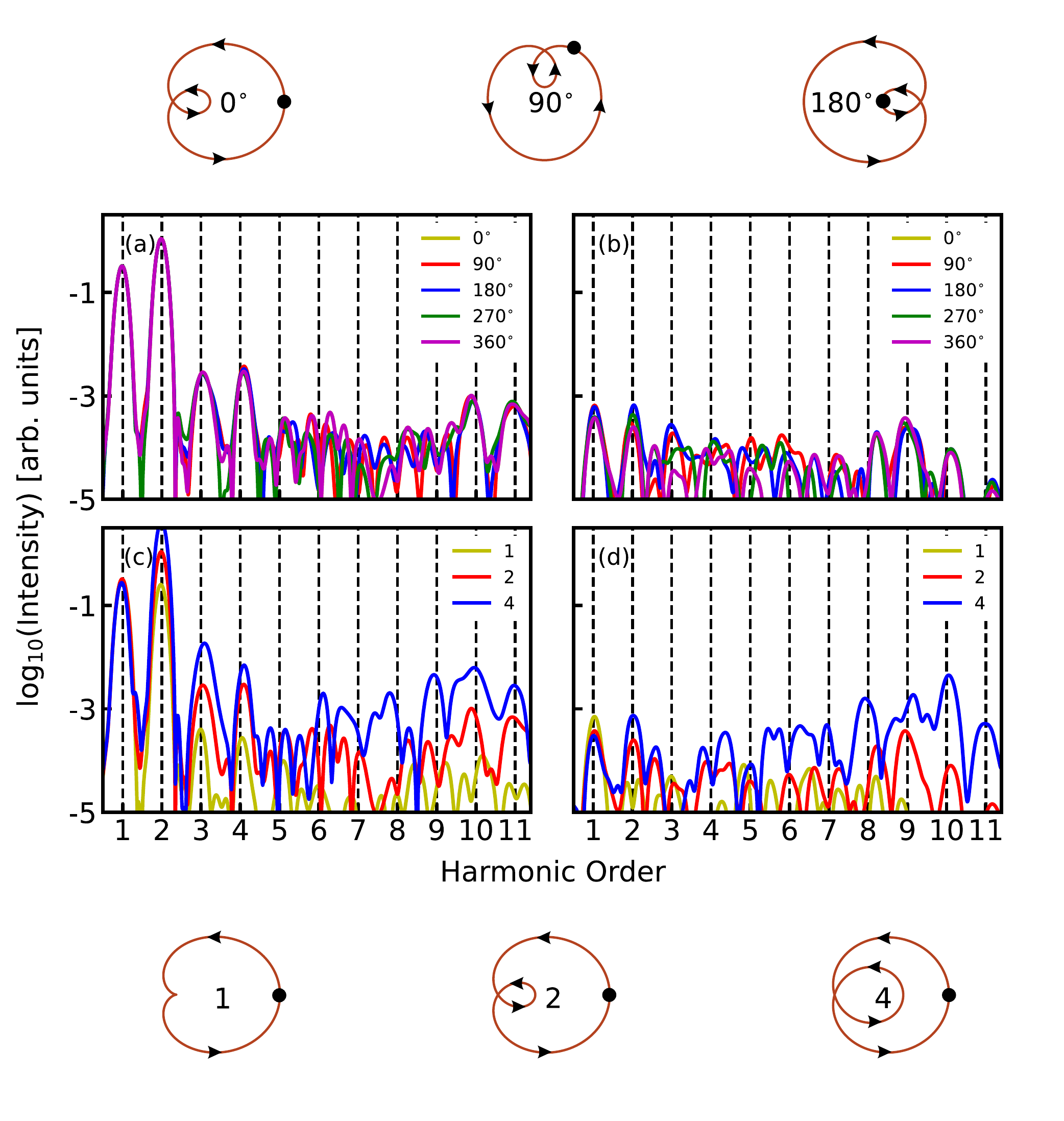}
\caption{High-harmonic spectra of graphene driven by  bicircular co-rotating $\omega$-2$\omega$ 
 laser pulses for different values of  (a)-(b) the subcycle phases $\phi$, 
and  (c)-(d) the relative electric field strengths $\mathsf{R}$. 
Left- and right-hand panels  show the contributions of left- and right-handed polarization to the harmonics, respectively. 
Lissajous figures of the resultant vector potential, associated with   bicircular co-rotating  $\omega$-2$\omega$ 
laser pulses, are shown in above and below the panels  for different values of $\phi$ and $\mathsf{R}$, respectively. 
The $x$ and $y$ components of the total vector potential as given in Eq.~(\ref{eq1}) 
are plotted omitting the envelope function along the $\textsf{X}$ and $\textsf{Y}$ axes for different values of $\phi$ and $\textsf{R}$ as mentioned in the inset. 
These inset  plots are aimed to display the symmetry of the total field corresponding to different phase and intensity relations between $\omega$ and $2\omega$ fields.
The black dot in each Lissajous figure corresponds to the start of the vector potential at $t =0 $.}
\label{phasevalue}
\end{figure}

At this juncture it is natural to wonder how robust are the features of the spectrum shown in Fig.~\ref{Spectra}(c) with respect to the variations in the subcycle phase $\phi$ and the relative ratio $\mathsf{R}$. 
To avoid  overlapping and have a better representation, 
we present  the harmonics with left- and right-handed circular polarization in the left- and right-hand panels of  
Fig.~\ref{phasevalue}, respectively.  
Figures~\ref{phasevalue}(a) and  \ref{phasevalue}(b) display  
the harmonic spectrum for different values of $\phi$. 
From the figure, the polarization state of the emitted harmonics is 
insensitive with respect to variations in $\phi$, 
and all the harmonics are left-handed circularly polarized in nature. To understand the insensitivity, let us 
look at the total electric field of the  co-rotating $\omega$-2$\omega$ 
scheme for different values of $\phi$ as shown above the panels. 
As evident from the Lissajous figures of the total field, the variation in $\phi$ results in the rotation of the resultant field. However,  the fields have no rotation 
symmetry.  Thus,  there are no significantly changes in the polarization and the  intensity of the emitted 
harmonics [see Figs.~\ref{phasevalue}(a) and \ref{phasevalue}(b)].  

\begin{figure}
\includegraphics[width=\linewidth]{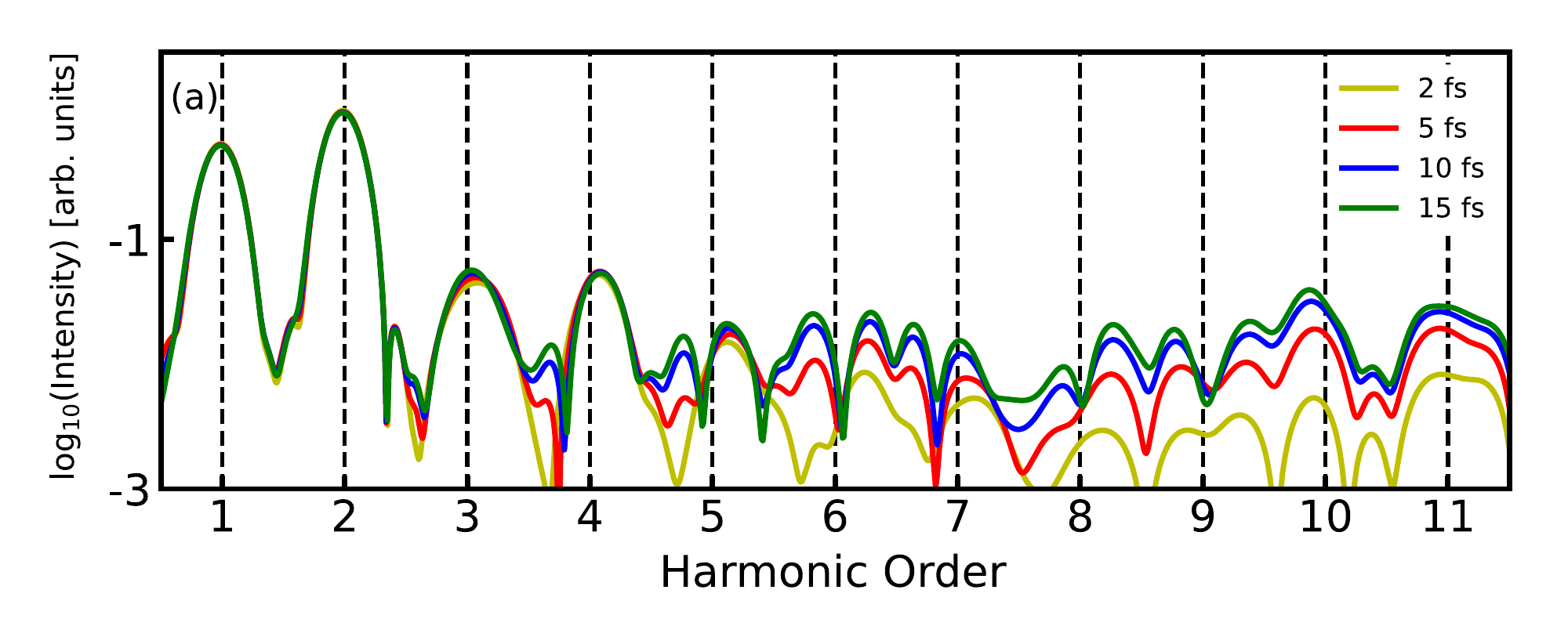}
\caption{High-harmonic spectrum of graphene generated by   bicircular co-rotating  $\omega$-2$\omega$ laser pulses for different values of dephasing time. 
Helicity of the emitted harmonics is the  same as that of the $\omega$ driving pulse, i.e., left-handed circularly polarized.
The laser parameters are the same as those used in Fig.~\ref{Spectra}(c).}
\label{dephasing}
\end{figure}

Now let us explore how the variation in the $\mathsf{R}$ value affects the harmonic spectra. 
The  peak intensity of the resultant field is maintained at 0.3 TW/cm$^{2}$ while varying the  $\mathsf{R}$ value. 
As expected, an increase  in the $\mathsf{R}$ value results in an increase of the harmonic yield  
[see Figs.~\ref{phasevalue}(c) and \ref{phasevalue}(d)].
However, the overall nature of the spectra does not change with respect to  $\mathsf{R}$, i.e., 
harmonics are still circularly polarized in the direction of the driving field. 
The reason for that can be attributed to the similar Lissajous figures of the resultant field for different 
$\mathsf{R}$ values as evident from the bottom panel. 
Thus, the investigation of Fig.~\ref{phasevalue} leads us to conclude  that the generation of  
circularly-polarized 
harmonics with same helicity via  the co-rotating $\omega$-2$\omega$ scheme does not require fine
tuning of subcycle phase $\phi$  or intensity ratio  $\mathsf{R}$ between the pulses. 
Therefore, the co-rotating scheme is robust in  generating circularly-polarized harmonics with the 
same helicity to all ordera. 

Before making a summary,  let us examine how the harmonic spectrum is sensitive to the dephasing time --  a phenomenological term accounting for the decoherence between electrons and holes within the  semiconductor-Bloch equations framework. 
Figure~\ref{dephasing} presents how the different values of
the dephasing time affect the harmonic spectrum corresponding to co-rotating $\omega$-2$\omega$  laser pulses.  
 As evident from the figure,  lower-order harmonics are insensitive to the dephasing time 
 as they are dominated by the intraband current ~\cite{mrudul2021high, taucer2017nonperturbative, heide2022probing}. 
On the other hand, higher-order harmonics are dominated by the interband current as 
their intensities are  boosted with an increase  in the dephasing time. 
The interband channel is prone to the coherence between the excited electrons and their corresponding holes~\cite{taucer2017nonperturbative}.  
Also, longer electron-hole trajectories are sensitive to the decoherence time as the corresponding excursion time allows for more scattering events during HHG~\cite{heide2022probing}. 
Thus, relatively smaller dephasing time means more scattering events for the same excursion time.  
Hence, the yield of higher harmonics will be suppressed  for smaller dephasing time~\cite{heide2022probing}.
Note that  the increase in the harmonic yield   with increase of dephasing time  does not alter the polarization properties of the emitted harmonics as evident from Fig.~\ref{dephasing}.

\section{Conclusion}  
In summary, we  explore the possibility of generating circularly-polarized high-harmonics with the same helicity in solids. To achieve this ambitious goal, 
we  harness  the no rotational symmetry of the tailored driving field consisting of two co-rotating circular pulses with frequencies $\omega$ and $2\omega$. 
The absence of  rotational symmetry of the driving field enforces no constraint on the helicity of the emitted harmonics. It is observed that the spectrum of monolayer graphene, driven by the bicircular co-rotating pulses, consists of circularly-polarized harmonics with the same helicity to all orders as that 
of the fundamental  $\omega$ pulse. 
The Lissajous figure of the total driving pulse exhibits no rotational symmetry for different subcycle phase differences between  $\omega$ and $2\omega$ pulses and different intensity ratios between the pulses.
Moreover, dephasing time, associated with the decoherence between electrons and holes, does not alter the 
nature of the harmonic spectrum. Thus, our approach for generating circularly-polarized high-harmonics with the 
same helicity is robust against any imperfection in the driving pulses. 
Furthermore, we anticipate that our 
approach will be applicable to other hexagonal two-dimensional materials, such as hexagonal boron nitride and molybdenum disulfide,  as our 
approach  is based on the absence of  rotational symmetry of the total system consisting of the total driving field and the two-dimensional material. 
However, unlike gapless graphene,  the interplay of the interband and intraband currents  
is different in gapped graphene~\cite{klemke2020role}, and therefore the value of the dephasing time  
will play a crucial role in determining the helicity of the emitted harmonics in gapped graphene. 
Thus, laser parameters need to be optimised  to obtain a desired control over the helicity of the emitted harmonics from gapped graphene. Also, based on the results presented in Ref.~\cite{heinrich2021chiral}, our approach could   be extended to bulk materials also.
Present work offers an avenue for detailed chiral-sensitive light-matter interactions on  
natural timescales.

\section*{Acknowledgements}
G. D. acknowledges support from Science and Engineering Research Board (SERB) India 
(Project No. MTR/2021/000138).    
\newpage

\bibliographystyle{revtex4-2}
\bibliography{solid_HHG}

\begin{thebibliography}{67}%
\makeatletter
\providecommand \@ifxundefined [1]{%
 \@ifx{#1\undefined}
}%
\providecommand \@ifnum [1]{%
 \ifnum #1\expandafter \@firstoftwo
 \else \expandafter \@secondoftwo
 \fi
}%
\providecommand \@ifx [1]{%
 \ifx #1\expandafter \@firstoftwo
 \else \expandafter \@secondoftwo
 \fi
}%
\providecommand \natexlab [1]{#1}%
\providecommand \enquote  [1]{``#1''}%
\providecommand \bibnamefont  [1]{#1}%
\providecommand \bibfnamefont [1]{#1}%
\providecommand \citenamefont [1]{#1}%
\providecommand \href@noop [0]{\@secondoftwo}%
\providecommand \href [0]{\begingroup \@sanitize@url \@href}%
\providecommand \@href[1]{\@@startlink{#1}\@@href}%
\providecommand \@@href[1]{\endgroup#1\@@endlink}%
\providecommand \@sanitize@url [0]{\catcode `\\12\catcode `\$12\catcode
  `\&12\catcode `\#12\catcode `\^12\catcode `\_12\catcode `\%12\relax}%
\providecommand \@@startlink[1]{}%
\providecommand \@@endlink[0]{}%
\providecommand \url  [0]{\begingroup\@sanitize@url \@url }%
\providecommand \@url [1]{\endgroup\@href {#1}{\urlprefix }}%
\providecommand \urlprefix  [0]{URL }%
\providecommand \Eprint [0]{\href }%
\providecommand \doibase [0]{https://doi.org/}%
\providecommand \selectlanguage [0]{\@gobble}%
\providecommand \bibinfo  [0]{\@secondoftwo}%
\providecommand \bibfield  [0]{\@secondoftwo}%
\providecommand \translation [1]{[#1]}%
\providecommand \BibitemOpen [0]{}%
\providecommand \bibitemStop [0]{}%
\providecommand \bibitemNoStop [0]{.\EOS\space}%
\providecommand \EOS [0]{\spacefactor3000\relax}%
\providecommand \BibitemShut  [1]{\csname bibitem#1\endcsname}%
\let\auto@bib@innerbib\@empty
\bibitem [{\citenamefont {Midorikawa}(2022)}]{midorikawa2022progress}%
  \BibitemOpen
  \bibfield  {author} {\bibinfo {author} {\bibfnamefont {K.}~\bibnamefont
  {Midorikawa}},\ }\bibfield  {title} {\bibinfo {title} {Progress on table-top
  isolated attosecond light sources},\ }\href@noop {} {\bibfield  {journal}
  {\bibinfo  {journal} {Nature Photonics}\ }\textbf {\bibinfo {volume} {16}},\
  \bibinfo {pages} {267} (\bibinfo {year} {2022})}\BibitemShut {NoStop}%
\bibitem [{\citenamefont {Krausz}\ and\ \citenamefont
  {Ivanov}(2009)}]{krausz2009attosecond}%
  \BibitemOpen
  \bibfield  {author} {\bibinfo {author} {\bibfnamefont {F.}~\bibnamefont
  {Krausz}}\ and\ \bibinfo {author} {\bibfnamefont {M.}~\bibnamefont
  {Ivanov}},\ }\bibfield  {title} {\bibinfo {title} {Attosecond physics},\
  }\href@noop {} {\bibfield  {journal} {\bibinfo  {journal} {Reviews of Modern
  Physics}\ }\textbf {\bibinfo {volume} {81}},\ \bibinfo {pages} {163}
  (\bibinfo {year} {2009})}\BibitemShut {NoStop}%
\bibitem [{\citenamefont {Bucksbaum}(2007)}]{bucksbaum2007}%
  \BibitemOpen
  \bibfield  {author} {\bibinfo {author} {\bibfnamefont {P.~H.}\ \bibnamefont
  {Bucksbaum}},\ }\bibfield  {title} {\bibinfo {title} {The future of
  attosecond spectroscopy},\ }\href@noop {} {\bibfield  {journal} {\bibinfo
  {journal} {Science}\ }\textbf {\bibinfo {volume} {317}},\ \bibinfo {pages}
  {766} (\bibinfo {year} {2007})}\BibitemShut {NoStop}%
\bibitem [{\citenamefont {Corkum}\ and\ \citenamefont
  {Krausz}(2007)}]{corkum2007}%
  \BibitemOpen
  \bibfield  {author} {\bibinfo {author} {\bibfnamefont {P.~B.}\ \bibnamefont
  {Corkum}}\ and\ \bibinfo {author} {\bibfnamefont {F.}~\bibnamefont
  {Krausz}},\ }\bibfield  {title} {\bibinfo {title} {Attosecond science},\
  }\href@noop {} {\bibfield  {journal} {\bibinfo  {journal} {Nature Physics}\
  }\textbf {\bibinfo {volume} {3}},\ \bibinfo {pages} {381} (\bibinfo {year}
  {2007})}\BibitemShut {NoStop}%
\bibitem [{\citenamefont {Dixit}\ \emph {et~al.}(2012)\citenamefont {Dixit},
  \citenamefont {Vendrell},\ and\ \citenamefont {Santra}}]{dixit2012}%
  \BibitemOpen
  \bibfield  {author} {\bibinfo {author} {\bibfnamefont {G.}~\bibnamefont
  {Dixit}}, \bibinfo {author} {\bibfnamefont {O.}~\bibnamefont {Vendrell}},\
  and\ \bibinfo {author} {\bibfnamefont {R.}~\bibnamefont {Santra}},\
  }\bibfield  {title} {\bibinfo {title} {Imaging electronic quantum motion with
  light},\ }\href@noop {} {\bibfield  {journal} {\bibinfo  {journal} {Proc.
  Natl. Acad. Sci. U.S.A.}\ }\textbf {\bibinfo {volume} {109}},\ \bibinfo
  {pages} {11636} (\bibinfo {year} {2012})}\BibitemShut {NoStop}%
\bibitem [{\citenamefont {Corkum}(1993)}]{corkum1993plasma}%
  \BibitemOpen
  \bibfield  {author} {\bibinfo {author} {\bibfnamefont {P.~B.}\ \bibnamefont
  {Corkum}},\ }\bibfield  {title} {\bibinfo {title} {Plasma perspective on
  strong field multiphoton ionization},\ }\href@noop {} {\bibfield  {journal}
  {\bibinfo  {journal} {Physical Review Letters}\ }\textbf {\bibinfo {volume}
  {71}},\ \bibinfo {pages} {1994} (\bibinfo {year} {1993})}\BibitemShut
  {NoStop}%
\bibitem [{\citenamefont {Neufeld}\ \emph {et~al.}(2019)\citenamefont
  {Neufeld}, \citenamefont {Podolsky},\ and\ \citenamefont
  {Cohen}}]{neufeld2019floquet}%
  \BibitemOpen
  \bibfield  {author} {\bibinfo {author} {\bibfnamefont {O.}~\bibnamefont
  {Neufeld}}, \bibinfo {author} {\bibfnamefont {D.}~\bibnamefont {Podolsky}},\
  and\ \bibinfo {author} {\bibfnamefont {O.}~\bibnamefont {Cohen}},\ }\bibfield
   {title} {\bibinfo {title} {Floquet group theory and its application to
  selection rules in harmonic generation},\ }\href@noop {} {\bibfield
  {journal} {\bibinfo  {journal} {Nature Communications}\ }\textbf {\bibinfo
  {volume} {10}},\ \bibinfo {pages} {1} (\bibinfo {year} {2019})}\BibitemShut
  {NoStop}%
\bibitem [{\citenamefont {Fleischer}\ \emph {et~al.}(2014)\citenamefont
  {Fleischer}, \citenamefont {Kfir}, \citenamefont {Diskin}, \citenamefont
  {Sidorenko},\ and\ \citenamefont {Cohen}}]{fleischer2014spin}%
  \BibitemOpen
  \bibfield  {author} {\bibinfo {author} {\bibfnamefont {A.}~\bibnamefont
  {Fleischer}}, \bibinfo {author} {\bibfnamefont {O.}~\bibnamefont {Kfir}},
  \bibinfo {author} {\bibfnamefont {T.}~\bibnamefont {Diskin}}, \bibinfo
  {author} {\bibfnamefont {P.}~\bibnamefont {Sidorenko}},\ and\ \bibinfo
  {author} {\bibfnamefont {O.}~\bibnamefont {Cohen}},\ }\bibfield  {title}
  {\bibinfo {title} {Spin angular momentum and tunable polarization in
  high-harmonic generation},\ }\href@noop {} {\bibfield  {journal} {\bibinfo
  {journal} {Nature Photonics}\ }\textbf {\bibinfo {volume} {8}},\ \bibinfo
  {pages} {543} (\bibinfo {year} {2014})}\BibitemShut {NoStop}%
\bibitem [{\citenamefont {Neufeld}\ and\ \citenamefont
  {Cohen}(2018)}]{neufeld2018optical}%
  \BibitemOpen
  \bibfield  {author} {\bibinfo {author} {\bibfnamefont {O.}~\bibnamefont
  {Neufeld}}\ and\ \bibinfo {author} {\bibfnamefont {O.}~\bibnamefont
  {Cohen}},\ }\bibfield  {title} {\bibinfo {title} {Optical chirality in
  nonlinear optics: Application to high harmonic generation},\ }\href@noop {}
  {\bibfield  {journal} {\bibinfo  {journal} {Physical Review Letters}\
  }\textbf {\bibinfo {volume} {120}},\ \bibinfo {pages} {133206} (\bibinfo
  {year} {2018})}\BibitemShut {NoStop}%
\bibitem [{\citenamefont {Dorney}\ \emph {et~al.}(2017)\citenamefont {Dorney},
  \citenamefont {Ellis}, \citenamefont {Hern{\'a}ndez-Garc{\'\i}a},
  \citenamefont {Hickstein}, \citenamefont {Mancuso}, \citenamefont {Brooks},
  \citenamefont {Fan}, \citenamefont {Fan}, \citenamefont {Zusin},
  \citenamefont {Gentry}, \citenamefont {Grychtol}, \citenamefont {Kapteyn},\
  and\ \citenamefont {Murnane}}]{dorney2017helicity}%
  \BibitemOpen
  \bibfield  {author} {\bibinfo {author} {\bibfnamefont {K.~M.}\ \bibnamefont
  {Dorney}}, \bibinfo {author} {\bibfnamefont {J.~L.}\ \bibnamefont {Ellis}},
  \bibinfo {author} {\bibfnamefont {C.}~\bibnamefont
  {Hern{\'a}ndez-Garc{\'\i}a}}, \bibinfo {author} {\bibfnamefont {D.~D.}\
  \bibnamefont {Hickstein}}, \bibinfo {author} {\bibfnamefont {C.~A.}\
  \bibnamefont {Mancuso}}, \bibinfo {author} {\bibfnamefont {N.}~\bibnamefont
  {Brooks}}, \bibinfo {author} {\bibfnamefont {T.}~\bibnamefont {Fan}},
  \bibinfo {author} {\bibfnamefont {G.}~\bibnamefont {Fan}}, \bibinfo {author}
  {\bibfnamefont {D.}~\bibnamefont {Zusin}}, \bibinfo {author} {\bibfnamefont
  {C.}~\bibnamefont {Gentry}}, \bibinfo {author} {\bibfnamefont
  {P.}~\bibnamefont {Grychtol}}, \bibinfo {author} {\bibfnamefont {H.~C.}\
  \bibnamefont {Kapteyn}},\ and\ \bibinfo {author} {\bibfnamefont {M.~M.}\
  \bibnamefont {Murnane}},\ }\bibfield  {title} {\bibinfo {title}
  {Helicity-selective enhancement and polarization control of attosecond high
  harmonic waveforms driven by bichromatic circularly polarized laser fields},\
  }\href@noop {} {\bibfield  {journal} {\bibinfo  {journal} {Physical Review
  Letters}\ }\textbf {\bibinfo {volume} {119}},\ \bibinfo {pages} {063201}
  (\bibinfo {year} {2017})}\BibitemShut {NoStop}%
\bibitem [{\citenamefont {Frolov}\ \emph {et~al.}(2018)\citenamefont {Frolov},
  \citenamefont {Manakov}, \citenamefont {Minina}, \citenamefont {Vvedenskii},
  \citenamefont {Silaev}, \citenamefont {Ivanov},\ and\ \citenamefont
  {Starace}}]{frolov2018control}%
  \BibitemOpen
  \bibfield  {author} {\bibinfo {author} {\bibfnamefont {M.~V.}\ \bibnamefont
  {Frolov}}, \bibinfo {author} {\bibfnamefont {N.~L.}\ \bibnamefont {Manakov}},
  \bibinfo {author} {\bibfnamefont {A.~A.}\ \bibnamefont {Minina}}, \bibinfo
  {author} {\bibfnamefont {N.~V.}\ \bibnamefont {Vvedenskii}}, \bibinfo
  {author} {\bibfnamefont {A.~A.}\ \bibnamefont {Silaev}}, \bibinfo {author}
  {\bibfnamefont {M.~Y.}\ \bibnamefont {Ivanov}},\ and\ \bibinfo {author}
  {\bibfnamefont {A.~F.}\ \bibnamefont {Starace}},\ }\bibfield  {title}
  {\bibinfo {title} {Control of harmonic generation by the time delay between
  two-color, bicircular few-cycle mid-ir laser pulses},\ }\href@noop {}
  {\bibfield  {journal} {\bibinfo  {journal} {Physical Review Letters}\
  }\textbf {\bibinfo {volume} {120}},\ \bibinfo {pages} {263203} (\bibinfo
  {year} {2018})}\BibitemShut {NoStop}%
\bibitem [{\citenamefont {Dixit}\ \emph {et~al.}(2018)\citenamefont {Dixit},
  \citenamefont {Jim{\'e}nez-Gal{\'a}n}, \citenamefont {Medi{\v{s}}auskas},\
  and\ \citenamefont {Ivanov}}]{dixit2018control}%
  \BibitemOpen
  \bibfield  {author} {\bibinfo {author} {\bibfnamefont {G.}~\bibnamefont
  {Dixit}}, \bibinfo {author} {\bibfnamefont {{\'A}.}~\bibnamefont
  {Jim{\'e}nez-Gal{\'a}n}}, \bibinfo {author} {\bibfnamefont {L.}~\bibnamefont
  {Medi{\v{s}}auskas}},\ and\ \bibinfo {author} {\bibfnamefont
  {M.}~\bibnamefont {Ivanov}},\ }\bibfield  {title} {\bibinfo {title} {Control
  of the helicity of high-order harmonic radiation using bichromatic circularly
  polarized laser fields},\ }\href@noop {} {\bibfield  {journal} {\bibinfo
  {journal} {Physical Review A}\ }\textbf {\bibinfo {volume} {98}},\ \bibinfo
  {pages} {053402} (\bibinfo {year} {2018})}\BibitemShut {NoStop}%
\bibitem [{\citenamefont {Rajpoot}\ \emph {et~al.}(2021)\citenamefont
  {Rajpoot}, \citenamefont {Holkundkar},\ and\ \citenamefont
  {Bandyopadhyay}}]{rajpoot2021polarization}%
  \BibitemOpen
  \bibfield  {author} {\bibinfo {author} {\bibfnamefont {R.}~\bibnamefont
  {Rajpoot}}, \bibinfo {author} {\bibfnamefont {A.~R.}\ \bibnamefont
  {Holkundkar}},\ and\ \bibinfo {author} {\bibfnamefont {J.~N.}\ \bibnamefont
  {Bandyopadhyay}},\ }\bibfield  {title} {\bibinfo {title} {Polarization
  control of attosecond pulses using bi-chromatic elliptically polarized
  laser},\ }\href@noop {} {\bibfield  {journal} {\bibinfo  {journal} {Journal
  of Physics B}\ }\textbf {\bibinfo {volume} {54}},\ \bibinfo {pages} {225401}
  (\bibinfo {year} {2021})}\BibitemShut {NoStop}%
\bibitem [{\citenamefont {Ansari}\ \emph {et~al.}(2021)\citenamefont {Ansari},
  \citenamefont {Hofmann}, \citenamefont {Medi{\v{s}}auskas}, \citenamefont
  {Lewenstein}, \citenamefont {Ciappina},\ and\ \citenamefont
  {Dixit}}]{ansari2021controlling}%
  \BibitemOpen
  \bibfield  {author} {\bibinfo {author} {\bibfnamefont {I.~N.}\ \bibnamefont
  {Ansari}}, \bibinfo {author} {\bibfnamefont {C.}~\bibnamefont {Hofmann}},
  \bibinfo {author} {\bibfnamefont {L.}~\bibnamefont {Medi{\v{s}}auskas}},
  \bibinfo {author} {\bibfnamefont {M.}~\bibnamefont {Lewenstein}}, \bibinfo
  {author} {\bibfnamefont {M.~F.}\ \bibnamefont {Ciappina}},\ and\ \bibinfo
  {author} {\bibfnamefont {G.}~\bibnamefont {Dixit}},\ }\bibfield  {title}
  {\bibinfo {title} {Controlling polarization of attosecond pulses with
  plasmonic-enhanced bichromatic counter-rotating circularly polarized
  fields},\ }\href@noop {} {\bibfield  {journal} {\bibinfo  {journal} {Physical
  Review A}\ }\textbf {\bibinfo {volume} {103}},\ \bibinfo {pages} {013104}
  (\bibinfo {year} {2021})}\BibitemShut {NoStop}%
\bibitem [{\citenamefont {Huang}\ \emph {et~al.}(2018)\citenamefont {Huang},
  \citenamefont {Hern{\'a}ndez-Garc{\'\i}a}, \citenamefont {Huang},
  \citenamefont {Huang}, \citenamefont {Lu}, \citenamefont {Rego},
  \citenamefont {Hickstein}, \citenamefont {Ellis}, \citenamefont
  {Jaron-Becker}, \citenamefont {Becker} \emph
  {et~al.}}]{huang2018polarization}%
  \BibitemOpen
  \bibfield  {author} {\bibinfo {author} {\bibfnamefont {P.~C.}\ \bibnamefont
  {Huang}}, \bibinfo {author} {\bibfnamefont {C.}~\bibnamefont
  {Hern{\'a}ndez-Garc{\'\i}a}}, \bibinfo {author} {\bibfnamefont {J.~T.}\
  \bibnamefont {Huang}}, \bibinfo {author} {\bibfnamefont {P.~Y.}\ \bibnamefont
  {Huang}}, \bibinfo {author} {\bibfnamefont {C.~H.}\ \bibnamefont {Lu}},
  \bibinfo {author} {\bibfnamefont {L.}~\bibnamefont {Rego}}, \bibinfo {author}
  {\bibfnamefont {D.~D.}\ \bibnamefont {Hickstein}}, \bibinfo {author}
  {\bibfnamefont {J.~L.}\ \bibnamefont {Ellis}}, \bibinfo {author}
  {\bibfnamefont {A.}~\bibnamefont {Jaron-Becker}}, \bibinfo {author}
  {\bibfnamefont {A.}~\bibnamefont {Becker}}, \emph {et~al.},\ }\bibfield
  {title} {\bibinfo {title} {Polarization control of isolated high-harmonic
  pulses},\ }\href@noop {} {\bibfield  {journal} {\bibinfo  {journal} {Nature
  Photonics}\ }\textbf {\bibinfo {volume} {12}},\ \bibinfo {pages} {349}
  (\bibinfo {year} {2018})}\BibitemShut {NoStop}%
\bibitem [{\citenamefont {Hickstein}\ \emph {et~al.}(2015)\citenamefont
  {Hickstein}, \citenamefont {Dollar}, \citenamefont {Grychtol}, \citenamefont
  {Ellis}, \citenamefont {Knut}, \citenamefont {Hern{\'a}ndez-Garc{\'\i}a},
  \citenamefont {Zusin}, \citenamefont {Gentry}, \citenamefont {Shaw},
  \citenamefont {Fan} \emph {et~al.}}]{hickstein2015non}%
  \BibitemOpen
  \bibfield  {author} {\bibinfo {author} {\bibfnamefont {D.~D.}\ \bibnamefont
  {Hickstein}}, \bibinfo {author} {\bibfnamefont {F.~J.}\ \bibnamefont
  {Dollar}}, \bibinfo {author} {\bibfnamefont {P.}~\bibnamefont {Grychtol}},
  \bibinfo {author} {\bibfnamefont {J.~L.}\ \bibnamefont {Ellis}}, \bibinfo
  {author} {\bibfnamefont {R.}~\bibnamefont {Knut}}, \bibinfo {author}
  {\bibfnamefont {C.}~\bibnamefont {Hern{\'a}ndez-Garc{\'\i}a}}, \bibinfo
  {author} {\bibfnamefont {D.}~\bibnamefont {Zusin}}, \bibinfo {author}
  {\bibfnamefont {C.}~\bibnamefont {Gentry}}, \bibinfo {author} {\bibfnamefont
  {J.~M.}\ \bibnamefont {Shaw}}, \bibinfo {author} {\bibfnamefont
  {T.}~\bibnamefont {Fan}}, \emph {et~al.},\ }\bibfield  {title} {\bibinfo
  {title} {Non-collinear generation of angularly isolated circularly polarized
  high harmonics},\ }\href@noop {} {\bibfield  {journal} {\bibinfo  {journal}
  {Nature Photonics}\ }\textbf {\bibinfo {volume} {9}},\ \bibinfo {pages} {743}
  (\bibinfo {year} {2015})}\BibitemShut {NoStop}%
\bibitem [{\citenamefont {Ferr{\'e}}\ \emph {et~al.}(2015)\citenamefont
  {Ferr{\'e}}, \citenamefont {Handschin}, \citenamefont {Dumergue},
  \citenamefont {Burgy}, \citenamefont {Comby}, \citenamefont {Descamps},
  \citenamefont {Fabre}, \citenamefont {Garcia}, \citenamefont {G{\'e}neaux},
  \citenamefont {Merceron}, \citenamefont {Mevel}, \citenamefont {Nahon},
  \citenamefont {Petit}, \citenamefont {Pons}, \citenamefont {Staedter},
  \citenamefont {Weber}, \citenamefont {Ruchon}, \citenamefont {Blanchet},\
  and\ \citenamefont {Mairesse}}]{ferre2015table}%
  \BibitemOpen
  \bibfield  {author} {\bibinfo {author} {\bibfnamefont {A.}~\bibnamefont
  {Ferr{\'e}}}, \bibinfo {author} {\bibfnamefont {C.}~\bibnamefont
  {Handschin}}, \bibinfo {author} {\bibfnamefont {M.}~\bibnamefont {Dumergue}},
  \bibinfo {author} {\bibfnamefont {F.}~\bibnamefont {Burgy}}, \bibinfo
  {author} {\bibfnamefont {A.}~\bibnamefont {Comby}}, \bibinfo {author}
  {\bibfnamefont {D.}~\bibnamefont {Descamps}}, \bibinfo {author}
  {\bibfnamefont {B.}~\bibnamefont {Fabre}}, \bibinfo {author} {\bibfnamefont
  {G.~A.}\ \bibnamefont {Garcia}}, \bibinfo {author} {\bibfnamefont
  {R.}~\bibnamefont {G{\'e}neaux}}, \bibinfo {author} {\bibfnamefont
  {L.}~\bibnamefont {Merceron}}, \bibinfo {author} {\bibfnamefont
  {E.}~\bibnamefont {Mevel}}, \bibinfo {author} {\bibfnamefont
  {L.}~\bibnamefont {Nahon}}, \bibinfo {author} {\bibfnamefont
  {S.}~\bibnamefont {Petit}}, \bibinfo {author} {\bibfnamefont
  {B.}~\bibnamefont {Pons}}, \bibinfo {author} {\bibfnamefont {D.}~\bibnamefont
  {Staedter}}, \bibinfo {author} {\bibfnamefont {S.}~\bibnamefont {Weber}},
  \bibinfo {author} {\bibfnamefont {T.}~\bibnamefont {Ruchon}}, \bibinfo
  {author} {\bibfnamefont {V.}~\bibnamefont {Blanchet}},\ and\ \bibinfo
  {author} {\bibfnamefont {Y.}~\bibnamefont {Mairesse}},\ }\bibfield  {title}
  {\bibinfo {title} {A table-top ultrashort light source in the extreme
  ultraviolet for circular dichroism experiments},\ }\href@noop {} {\bibfield
  {journal} {\bibinfo  {journal} {Nature Photonics}\ }\textbf {\bibinfo
  {volume} {9}},\ \bibinfo {pages} {93} (\bibinfo {year} {2015})}\BibitemShut
  {NoStop}%
\bibitem [{\citenamefont {Cireasa}\ \emph {et~al.}(2015)\citenamefont
  {Cireasa}, \citenamefont {Boguslavskiy}, \citenamefont {Pons}, \citenamefont
  {Wong}, \citenamefont {Descamps}, \citenamefont {Petit}, \citenamefont {Ruf},
  \citenamefont {Thir{\'e}}, \citenamefont {Ferr{\'e}}, \citenamefont {Suarez},
  \citenamefont {Higuet}, \citenamefont {Schmidt}, \citenamefont {Alharbi},
  \citenamefont {Legare}, \citenamefont {Blanchet}, \citenamefont {Fabre},
  \citenamefont {Patchkovskii}, \citenamefont {Smirnova}, \citenamefont
  {Mairesse}, ,\ and\ \citenamefont {Bhardwaj}}]{cireasa2015probing}%
  \BibitemOpen
  \bibfield  {author} {\bibinfo {author} {\bibfnamefont {R.}~\bibnamefont
  {Cireasa}}, \bibinfo {author} {\bibfnamefont {A.~E.}\ \bibnamefont
  {Boguslavskiy}}, \bibinfo {author} {\bibfnamefont {B.}~\bibnamefont {Pons}},
  \bibinfo {author} {\bibfnamefont {M.~C.~H.}\ \bibnamefont {Wong}}, \bibinfo
  {author} {\bibfnamefont {D.}~\bibnamefont {Descamps}}, \bibinfo {author}
  {\bibfnamefont {S.}~\bibnamefont {Petit}}, \bibinfo {author} {\bibfnamefont
  {H.}~\bibnamefont {Ruf}}, \bibinfo {author} {\bibfnamefont {N.}~\bibnamefont
  {Thir{\'e}}}, \bibinfo {author} {\bibfnamefont {A.}~\bibnamefont
  {Ferr{\'e}}}, \bibinfo {author} {\bibfnamefont {J.}~\bibnamefont {Suarez}},
  \bibinfo {author} {\bibfnamefont {J.}~\bibnamefont {Higuet}}, \bibinfo
  {author} {\bibfnamefont {B.~E.}\ \bibnamefont {Schmidt}}, \bibinfo {author}
  {\bibfnamefont {A.~F.}\ \bibnamefont {Alharbi}}, \bibinfo {author}
  {\bibfnamefont {F.}~\bibnamefont {Legare}}, \bibinfo {author} {\bibfnamefont
  {V.}~\bibnamefont {Blanchet}}, \bibinfo {author} {\bibfnamefont
  {B.}~\bibnamefont {Fabre}}, \bibinfo {author} {\bibfnamefont
  {S.}~\bibnamefont {Patchkovskii}}, \bibinfo {author} {\bibfnamefont
  {O.}~\bibnamefont {Smirnova}}, \bibinfo {author} {\bibfnamefont
  {Y.}~\bibnamefont {Mairesse}}, ,\ and\ \bibinfo {author} {\bibfnamefont
  {V.~R.}\ \bibnamefont {Bhardwaj}},\ }\bibfield  {title} {\bibinfo {title}
  {Probing molecular chirality on a sub-femtosecond timescale},\ }\href@noop {}
  {\bibfield  {journal} {\bibinfo  {journal} {Nature Physics}\ }\textbf
  {\bibinfo {volume} {11}},\ \bibinfo {pages} {654} (\bibinfo {year}
  {2015})}\BibitemShut {NoStop}%
\bibitem [{\citenamefont {Kfir}\ \emph {et~al.}(2015)\citenamefont {Kfir},
  \citenamefont {Grychtol}, \citenamefont {Turgut}, \citenamefont {Knut},
  \citenamefont {Zusin}, \citenamefont {Popmintchev}, \citenamefont
  {Popmintchev}, \citenamefont {Nembach}, \citenamefont {Shaw}, \citenamefont
  {Fleischer}, \citenamefont {Kapteyn}, \citenamefont {Murnane},\ and\
  \citenamefont {Cohen}}]{kfir2015generation}%
  \BibitemOpen
  \bibfield  {author} {\bibinfo {author} {\bibfnamefont {O.}~\bibnamefont
  {Kfir}}, \bibinfo {author} {\bibfnamefont {P.}~\bibnamefont {Grychtol}},
  \bibinfo {author} {\bibfnamefont {E.}~\bibnamefont {Turgut}}, \bibinfo
  {author} {\bibfnamefont {R.}~\bibnamefont {Knut}}, \bibinfo {author}
  {\bibfnamefont {D.}~\bibnamefont {Zusin}}, \bibinfo {author} {\bibfnamefont
  {D.}~\bibnamefont {Popmintchev}}, \bibinfo {author} {\bibfnamefont
  {T.}~\bibnamefont {Popmintchev}}, \bibinfo {author} {\bibfnamefont
  {H.}~\bibnamefont {Nembach}}, \bibinfo {author} {\bibfnamefont {J.~M.}\
  \bibnamefont {Shaw}}, \bibinfo {author} {\bibfnamefont {A.}~\bibnamefont
  {Fleischer}}, \bibinfo {author} {\bibfnamefont {H.}~\bibnamefont {Kapteyn}},
  \bibinfo {author} {\bibfnamefont {M.}~\bibnamefont {Murnane}},\ and\ \bibinfo
  {author} {\bibfnamefont {O.}~\bibnamefont {Cohen}},\ }\bibfield  {title}
  {\bibinfo {title} {Generation of bright phase-matched circularly-polarized
  extreme ultraviolet high harmonics},\ }\href@noop {} {\bibfield  {journal}
  {\bibinfo  {journal} {Nature Photonics}\ }\textbf {\bibinfo {volume} {9}},\
  \bibinfo {pages} {99} (\bibinfo {year} {2015})}\BibitemShut {NoStop}%
\bibitem [{\citenamefont {Boeglin}\ \emph {et~al.}(2010)\citenamefont
  {Boeglin}, \citenamefont {Beaurepaire}, \citenamefont {Halt{\'e}},
  \citenamefont {L{\'o}pez-Flores}, \citenamefont {Stamm}, \citenamefont
  {Pontius}, \citenamefont {D{\"u}rr},\ and\ \citenamefont
  {Bigot}}]{boeglin2010distinguishing}%
  \BibitemOpen
  \bibfield  {author} {\bibinfo {author} {\bibfnamefont {C.}~\bibnamefont
  {Boeglin}}, \bibinfo {author} {\bibfnamefont {E.}~\bibnamefont
  {Beaurepaire}}, \bibinfo {author} {\bibfnamefont {V.}~\bibnamefont
  {Halt{\'e}}}, \bibinfo {author} {\bibfnamefont {V.}~\bibnamefont
  {L{\'o}pez-Flores}}, \bibinfo {author} {\bibfnamefont {C.}~\bibnamefont
  {Stamm}}, \bibinfo {author} {\bibfnamefont {N.}~\bibnamefont {Pontius}},
  \bibinfo {author} {\bibfnamefont {H.~A.}\ \bibnamefont {D{\"u}rr}},\ and\
  \bibinfo {author} {\bibfnamefont {J.~Y.}\ \bibnamefont {Bigot}},\ }\bibfield
  {title} {\bibinfo {title} {Distinguishing the ultrafast dynamics of spin and
  orbital moments in solids},\ }\href@noop {} {\bibfield  {journal} {\bibinfo
  {journal} {Nature}\ }\textbf {\bibinfo {volume} {465}},\ \bibinfo {pages}
  {458} (\bibinfo {year} {2010})}\BibitemShut {NoStop}%
\bibitem [{\citenamefont {Radu}\ \emph {et~al.}(2011)\citenamefont {Radu},
  \citenamefont {Vahaplar}, \citenamefont {Stamm}, \citenamefont {Kachel},
  \citenamefont {Pontius}, \citenamefont {D{\"u}rr}, \citenamefont {Ostler},
  \citenamefont {Barker}, \citenamefont {Evans}, \citenamefont {Chantrell},
  \citenamefont {Tsukamoto}, \citenamefont {Itoh}, \citenamefont {Kirilyuk},
  \citenamefont {Rasing},\ and\ \citenamefont {Kimel}}]{radu2011transient}%
  \BibitemOpen
  \bibfield  {author} {\bibinfo {author} {\bibfnamefont {I.}~\bibnamefont
  {Radu}}, \bibinfo {author} {\bibfnamefont {K.}~\bibnamefont {Vahaplar}},
  \bibinfo {author} {\bibfnamefont {C.}~\bibnamefont {Stamm}}, \bibinfo
  {author} {\bibfnamefont {T.}~\bibnamefont {Kachel}}, \bibinfo {author}
  {\bibfnamefont {N.}~\bibnamefont {Pontius}}, \bibinfo {author} {\bibfnamefont
  {H.~A.}\ \bibnamefont {D{\"u}rr}}, \bibinfo {author} {\bibfnamefont {T.~A.}\
  \bibnamefont {Ostler}}, \bibinfo {author} {\bibfnamefont {J.}~\bibnamefont
  {Barker}}, \bibinfo {author} {\bibfnamefont {R.~F.~L.}\ \bibnamefont
  {Evans}}, \bibinfo {author} {\bibfnamefont {R.~W.}\ \bibnamefont
  {Chantrell}}, \bibinfo {author} {\bibfnamefont {A.}~\bibnamefont
  {Tsukamoto}}, \bibinfo {author} {\bibfnamefont {A.}~\bibnamefont {Itoh}},
  \bibinfo {author} {\bibfnamefont {A.}~\bibnamefont {Kirilyuk}}, \bibinfo
  {author} {\bibfnamefont {T.}~\bibnamefont {Rasing}},\ and\ \bibinfo {author}
  {\bibfnamefont {A.~V.}\ \bibnamefont {Kimel}},\ }\bibfield  {title} {\bibinfo
  {title} {Transient ferromagnetic-like state mediating ultrafast reversal of
  antiferromagnetically coupled spins},\ }\href@noop {} {\bibfield  {journal}
  {\bibinfo  {journal} {Nature}\ }\textbf {\bibinfo {volume} {472}},\ \bibinfo
  {pages} {205} (\bibinfo {year} {2011})}\BibitemShut {NoStop}%
\bibitem [{\citenamefont {Goulielmakis}\ and\ \citenamefont
  {Brabec}(2022)}]{goulielmakis2022high}%
  \BibitemOpen
  \bibfield  {author} {\bibinfo {author} {\bibfnamefont {E.}~\bibnamefont
  {Goulielmakis}}\ and\ \bibinfo {author} {\bibfnamefont {T.}~\bibnamefont
  {Brabec}},\ }\bibfield  {title} {\bibinfo {title} {High harmonic generation
  in condensed matter},\ }\href@noop {} {\bibfield  {journal} {\bibinfo
  {journal} {Nature Photonics}\ }\textbf {\bibinfo {volume} {16}},\ \bibinfo
  {pages} {411} (\bibinfo {year} {2022})}\BibitemShut {NoStop}%
\bibitem [{\citenamefont {Ghimire}\ and\ \citenamefont
  {Reis}(2019)}]{ghimire2019}%
  \BibitemOpen
  \bibfield  {author} {\bibinfo {author} {\bibfnamefont {S.}~\bibnamefont
  {Ghimire}}\ and\ \bibinfo {author} {\bibfnamefont {D.~A.}\ \bibnamefont
  {Reis}},\ }\bibfield  {title} {\bibinfo {title} {High-harmonic generation
  from solids},\ }\href@noop {} {\bibfield  {journal} {\bibinfo  {journal}
  {Nature Physics}\ }\textbf {\bibinfo {volume} {15}},\ \bibinfo {pages} {10}
  (\bibinfo {year} {2019})}\BibitemShut {NoStop}%
\bibitem [{\citenamefont {Bharti}\ \emph {et~al.}(2022)\citenamefont {Bharti},
  \citenamefont {Mrudul},\ and\ \citenamefont {Dixit}}]{bharti2022high}%
  \BibitemOpen
  \bibfield  {author} {\bibinfo {author} {\bibfnamefont {A.}~\bibnamefont
  {Bharti}}, \bibinfo {author} {\bibfnamefont {M.}~\bibnamefont {Mrudul}},\
  and\ \bibinfo {author} {\bibfnamefont {G.}~\bibnamefont {Dixit}},\ }\bibfield
   {title} {\bibinfo {title} {High-harmonic spectroscopy of light-driven
  nonlinear anisotropic anomalous hall effect in a weyl semimetal},\
  }\href@noop {} {\bibfield  {journal} {\bibinfo  {journal} {Physical Review
  B}\ }\textbf {\bibinfo {volume} {105}},\ \bibinfo {pages} {155140} (\bibinfo
  {year} {2022})}\BibitemShut {NoStop}%
\bibitem [{\citenamefont {Pattanayak}\ \emph {et~al.}(2022)\citenamefont
  {Pattanayak}, \citenamefont {Pujari},\ and\ \citenamefont
  {Dixit}}]{pattanayak2022role}%
  \BibitemOpen
  \bibfield  {author} {\bibinfo {author} {\bibfnamefont {A.}~\bibnamefont
  {Pattanayak}}, \bibinfo {author} {\bibfnamefont {S.}~\bibnamefont {Pujari}},\
  and\ \bibinfo {author} {\bibfnamefont {G.}~\bibnamefont {Dixit}},\ }\bibfield
   {title} {\bibinfo {title} {Role of majorana fermions in high-harmonic
  generation from kitaev chain},\ }\href@noop {} {\bibfield  {journal}
  {\bibinfo  {journal} {Scientific Reports}\ }\textbf {\bibinfo {volume}
  {12}},\ \bibinfo {pages} {1} (\bibinfo {year} {2022})}\BibitemShut {NoStop}%
\bibitem [{\citenamefont {Zaks}\ \emph {et~al.}(2012)\citenamefont {Zaks},
  \citenamefont {Liu},\ and\ \citenamefont {Sherwin}}]{zaks2012experimental}%
  \BibitemOpen
  \bibfield  {author} {\bibinfo {author} {\bibfnamefont {B.}~\bibnamefont
  {Zaks}}, \bibinfo {author} {\bibfnamefont {R.-B.}\ \bibnamefont {Liu}},\ and\
  \bibinfo {author} {\bibfnamefont {M.~S.}\ \bibnamefont {Sherwin}},\
  }\bibfield  {title} {\bibinfo {title} {Experimental observation of
  electron--hole recollisions},\ }\href@noop {} {\bibfield  {journal} {\bibinfo
   {journal} {Nature}\ }\textbf {\bibinfo {volume} {483}},\ \bibinfo {pages}
  {580} (\bibinfo {year} {2012})}\BibitemShut {NoStop}%
\bibitem [{\citenamefont {Luu}\ \emph {et~al.}(2015)\citenamefont {Luu},
  \citenamefont {Garg}, \citenamefont {Kruchinin}, \citenamefont {Moulet},
  \citenamefont {Hassan},\ and\ \citenamefont {Goulielmakis}}]{luu2015extreme}%
  \BibitemOpen
  \bibfield  {author} {\bibinfo {author} {\bibfnamefont {T.~T.}\ \bibnamefont
  {Luu}}, \bibinfo {author} {\bibfnamefont {M.}~\bibnamefont {Garg}}, \bibinfo
  {author} {\bibfnamefont {S.~Y.}\ \bibnamefont {Kruchinin}}, \bibinfo {author}
  {\bibfnamefont {A.}~\bibnamefont {Moulet}}, \bibinfo {author} {\bibfnamefont
  {M.~T.}\ \bibnamefont {Hassan}},\ and\ \bibinfo {author} {\bibfnamefont
  {E.}~\bibnamefont {Goulielmakis}},\ }\bibfield  {title} {\bibinfo {title}
  {Extreme ultraviolet high-harmonic spectroscopy of solids},\ }\href@noop {}
  {\bibfield  {journal} {\bibinfo  {journal} {Nature}\ }\textbf {\bibinfo
  {volume} {521}},\ \bibinfo {pages} {498} (\bibinfo {year}
  {2015})}\BibitemShut {NoStop}%
\bibitem [{\citenamefont {Schubert}\ \emph {et~al.}(2014)\citenamefont
  {Schubert}, \citenamefont {Hohenleutner}, \citenamefont {Langer},
  \citenamefont {Urbanek}, \citenamefont {Lange}, \citenamefont {Huttner},
  \citenamefont {Golde}, \citenamefont {Meier}, \citenamefont {Kira},
  \citenamefont {Koch},\ and\ \citenamefont {Huber}}]{schubert2014sub}%
  \BibitemOpen
  \bibfield  {author} {\bibinfo {author} {\bibfnamefont {O.}~\bibnamefont
  {Schubert}}, \bibinfo {author} {\bibfnamefont {M.}~\bibnamefont
  {Hohenleutner}}, \bibinfo {author} {\bibfnamefont {F.}~\bibnamefont
  {Langer}}, \bibinfo {author} {\bibfnamefont {B.}~\bibnamefont {Urbanek}},
  \bibinfo {author} {\bibfnamefont {C.}~\bibnamefont {Lange}}, \bibinfo
  {author} {\bibfnamefont {U.}~\bibnamefont {Huttner}}, \bibinfo {author}
  {\bibfnamefont {D.}~\bibnamefont {Golde}}, \bibinfo {author} {\bibfnamefont
  {T.}~\bibnamefont {Meier}}, \bibinfo {author} {\bibfnamefont
  {M.}~\bibnamefont {Kira}}, \bibinfo {author} {\bibfnamefont {S.~W.}\
  \bibnamefont {Koch}},\ and\ \bibinfo {author} {\bibfnamefont
  {R.}~\bibnamefont {Huber}},\ }\bibfield  {title} {\bibinfo {title} {Sub-cycle
  control of terahertz high-harmonic generation by dynamical bloch
  oscillations},\ }\href@noop {} {\bibfield  {journal} {\bibinfo  {journal}
  {Nature Photonics}\ }\textbf {\bibinfo {volume} {8}},\ \bibinfo {pages} {119}
  (\bibinfo {year} {2014})}\BibitemShut {NoStop}%
\bibitem [{\citenamefont {Hohenleutner}\ \emph {et~al.}(2015)\citenamefont
  {Hohenleutner}, \citenamefont {Langer}, \citenamefont {Schubert},
  \citenamefont {Knorr}, \citenamefont {Huttner}, \citenamefont {Koch},
  \citenamefont {Kira},\ and\ \citenamefont {Huber}}]{hohenleutner2015real}%
  \BibitemOpen
  \bibfield  {author} {\bibinfo {author} {\bibfnamefont {M.}~\bibnamefont
  {Hohenleutner}}, \bibinfo {author} {\bibfnamefont {F.}~\bibnamefont
  {Langer}}, \bibinfo {author} {\bibfnamefont {O.}~\bibnamefont {Schubert}},
  \bibinfo {author} {\bibfnamefont {M.}~\bibnamefont {Knorr}}, \bibinfo
  {author} {\bibfnamefont {U.}~\bibnamefont {Huttner}}, \bibinfo {author}
  {\bibfnamefont {S.~W.}\ \bibnamefont {Koch}}, \bibinfo {author}
  {\bibfnamefont {M.}~\bibnamefont {Kira}},\ and\ \bibinfo {author}
  {\bibfnamefont {R.}~\bibnamefont {Huber}},\ }\bibfield  {title} {\bibinfo
  {title} {Real-time observation of interfering crystal electrons in
  high-harmonic generation},\ }\href@noop {} {\bibfield  {journal} {\bibinfo
  {journal} {Nature}\ }\textbf {\bibinfo {volume} {523}},\ \bibinfo {pages}
  {572} (\bibinfo {year} {2015})}\BibitemShut {NoStop}%
\bibitem [{\citenamefont {Pattanayak}\ \emph {et~al.}(2020)\citenamefont
  {Pattanayak}, \citenamefont {Mrudul},\ and\ \citenamefont
  {Dixit}}]{pattanayak2020influence}%
  \BibitemOpen
  \bibfield  {author} {\bibinfo {author} {\bibfnamefont {A.}~\bibnamefont
  {Pattanayak}}, \bibinfo {author} {\bibfnamefont {M.~S.}\ \bibnamefont
  {Mrudul}},\ and\ \bibinfo {author} {\bibfnamefont {G.}~\bibnamefont
  {Dixit}},\ }\bibfield  {title} {\bibinfo {title} {Influence of vacancy
  defects in solid high-order harmonic generation},\ }\href@noop {} {\bibfield
  {journal} {\bibinfo  {journal} {Physical Review A}\ }\textbf {\bibinfo
  {volume} {101}},\ \bibinfo {pages} {013404} (\bibinfo {year}
  {2020})}\BibitemShut {NoStop}%
\bibitem [{\citenamefont {Vampa}\ \emph
  {et~al.}(2015{\natexlab{a}})\citenamefont {Vampa}, \citenamefont {Hammond},
  \citenamefont {Thir{\'e}}, \citenamefont {Schmidt}, \citenamefont
  {L{\'e}gar{\'e}}, \citenamefont {McDonald}, \citenamefont {Brabec},\ and\
  \citenamefont {Corkum}}]{vampa2015linking}%
  \BibitemOpen
  \bibfield  {author} {\bibinfo {author} {\bibfnamefont {G.}~\bibnamefont
  {Vampa}}, \bibinfo {author} {\bibfnamefont {T.~J.}\ \bibnamefont {Hammond}},
  \bibinfo {author} {\bibfnamefont {N.}~\bibnamefont {Thir{\'e}}}, \bibinfo
  {author} {\bibfnamefont {B.~E.}\ \bibnamefont {Schmidt}}, \bibinfo {author}
  {\bibfnamefont {F.}~\bibnamefont {L{\'e}gar{\'e}}}, \bibinfo {author}
  {\bibfnamefont {C.~R.}\ \bibnamefont {McDonald}}, \bibinfo {author}
  {\bibfnamefont {T.}~\bibnamefont {Brabec}},\ and\ \bibinfo {author}
  {\bibfnamefont {P.~B.}\ \bibnamefont {Corkum}},\ }\bibfield  {title}
  {\bibinfo {title} {Linking high harmonics from gases and solids},\
  }\href@noop {} {\bibfield  {journal} {\bibinfo  {journal} {Nature}\ }\textbf
  {\bibinfo {volume} {522}},\ \bibinfo {pages} {462} (\bibinfo {year}
  {2015}{\natexlab{a}})}\BibitemShut {NoStop}%
\bibitem [{\citenamefont {Vampa}\ \emph
  {et~al.}(2015{\natexlab{b}})\citenamefont {Vampa}, \citenamefont {Hammond},
  \citenamefont {Thir{\'e}}, \citenamefont {Schmidt}, \citenamefont
  {L{\'e}gar{\'e}}, \citenamefont {McDonald}, \citenamefont {Brabec},
  \citenamefont {Klug},\ and\ \citenamefont {Corkum}}]{vampa2015all}%
  \BibitemOpen
  \bibfield  {author} {\bibinfo {author} {\bibfnamefont {G.}~\bibnamefont
  {Vampa}}, \bibinfo {author} {\bibfnamefont {T.~J.}\ \bibnamefont {Hammond}},
  \bibinfo {author} {\bibfnamefont {N.}~\bibnamefont {Thir{\'e}}}, \bibinfo
  {author} {\bibfnamefont {B.~E.}\ \bibnamefont {Schmidt}}, \bibinfo {author}
  {\bibfnamefont {F.}~\bibnamefont {L{\'e}gar{\'e}}}, \bibinfo {author}
  {\bibfnamefont {C.~R.}\ \bibnamefont {McDonald}}, \bibinfo {author}
  {\bibfnamefont {T.}~\bibnamefont {Brabec}}, \bibinfo {author} {\bibfnamefont
  {D.~D.}\ \bibnamefont {Klug}},\ and\ \bibinfo {author} {\bibfnamefont
  {P.~B.}\ \bibnamefont {Corkum}},\ }\bibfield  {title} {\bibinfo {title}
  {All-optical reconstruction of crystal band structure},\ }\href@noop {}
  {\bibfield  {journal} {\bibinfo  {journal} {Physical Review Letters}\
  }\textbf {\bibinfo {volume} {115}},\ \bibinfo {pages} {193603} (\bibinfo
  {year} {2015}{\natexlab{b}})}\BibitemShut {NoStop}%
\bibitem [{\citenamefont {Langer}\ \emph {et~al.}(2018)\citenamefont {Langer},
  \citenamefont {Schmid}, \citenamefont {Schlauderer}, \citenamefont {Gmitra},
  \citenamefont {Fabian}, \citenamefont {Nagler}, \citenamefont {Sch{\"u}ller},
  \citenamefont {Korn}, \citenamefont {Hawkins}, \citenamefont {Steiner} \emph
  {et~al.}}]{langer2018lightwave}%
  \BibitemOpen
  \bibfield  {author} {\bibinfo {author} {\bibfnamefont {F.}~\bibnamefont
  {Langer}}, \bibinfo {author} {\bibfnamefont {C.~P.}\ \bibnamefont {Schmid}},
  \bibinfo {author} {\bibfnamefont {S.}~\bibnamefont {Schlauderer}}, \bibinfo
  {author} {\bibfnamefont {M.}~\bibnamefont {Gmitra}}, \bibinfo {author}
  {\bibfnamefont {J.}~\bibnamefont {Fabian}}, \bibinfo {author} {\bibfnamefont
  {P.}~\bibnamefont {Nagler}}, \bibinfo {author} {\bibfnamefont
  {C.}~\bibnamefont {Sch{\"u}ller}}, \bibinfo {author} {\bibfnamefont
  {T.}~\bibnamefont {Korn}}, \bibinfo {author} {\bibfnamefont {P.}~\bibnamefont
  {Hawkins}}, \bibinfo {author} {\bibfnamefont {J.}~\bibnamefont {Steiner}},
  \emph {et~al.},\ }\bibfield  {title} {\bibinfo {title} {Lightwave
  valleytronics in a monolayer of tungsten diselenide},\ }\href@noop {}
  {\bibfield  {journal} {\bibinfo  {journal} {Nature}\ }\textbf {\bibinfo
  {volume} {557}},\ \bibinfo {pages} {76} (\bibinfo {year} {2018})}\BibitemShut
  {NoStop}%
\bibitem [{\citenamefont {Luu}\ and\ \citenamefont
  {W{\"o}rner}(2018)}]{luu2018measurement}%
  \BibitemOpen
  \bibfield  {author} {\bibinfo {author} {\bibfnamefont {T.~T.}\ \bibnamefont
  {Luu}}\ and\ \bibinfo {author} {\bibfnamefont {H.~J.}\ \bibnamefont
  {W{\"o}rner}},\ }\bibfield  {title} {\bibinfo {title} {Measurement of the
  berry curvature of solids using high-harmonic spectroscopy},\ }\href@noop {}
  {\bibfield  {journal} {\bibinfo  {journal} {Nature Communications}\ }\textbf
  {\bibinfo {volume} {9}},\ \bibinfo {pages} {1} (\bibinfo {year}
  {2018})}\BibitemShut {NoStop}%
\bibitem [{\citenamefont {Banks}\ \emph {et~al.}(2017)\citenamefont {Banks},
  \citenamefont {Wu}, \citenamefont {Valovcin}, \citenamefont {Mack},
  \citenamefont {Gossard}, \citenamefont {Pfeiffer}, \citenamefont {Liu},\ and\
  \citenamefont {Sherwin}}]{banks2017dynamical}%
  \BibitemOpen
  \bibfield  {author} {\bibinfo {author} {\bibfnamefont {H.~B.}\ \bibnamefont
  {Banks}}, \bibinfo {author} {\bibfnamefont {Q.}~\bibnamefont {Wu}}, \bibinfo
  {author} {\bibfnamefont {D.~C.}\ \bibnamefont {Valovcin}}, \bibinfo {author}
  {\bibfnamefont {S.}~\bibnamefont {Mack}}, \bibinfo {author} {\bibfnamefont
  {A.~C.}\ \bibnamefont {Gossard}}, \bibinfo {author} {\bibfnamefont
  {L.}~\bibnamefont {Pfeiffer}}, \bibinfo {author} {\bibfnamefont {R.-B.}\
  \bibnamefont {Liu}},\ and\ \bibinfo {author} {\bibfnamefont {M.~S.}\
  \bibnamefont {Sherwin}},\ }\bibfield  {title} {\bibinfo {title} {Dynamical
  birefringence: electron-hole recollisions as probes of berry curvature},\
  }\href@noop {} {\bibfield  {journal} {\bibinfo  {journal} {Physical Review
  X}\ }\textbf {\bibinfo {volume} {7}},\ \bibinfo {pages} {041042} (\bibinfo
  {year} {2017})}\BibitemShut {NoStop}%
\bibitem [{\citenamefont {Mrudul}\ \emph {et~al.}(2019)\citenamefont {Mrudul},
  \citenamefont {Pattanayak}, \citenamefont {Ivanov},\ and\ \citenamefont
  {Dixit}}]{pattanayak2019direct}%
  \BibitemOpen
  \bibfield  {author} {\bibinfo {author} {\bibfnamefont {M.~S.}\ \bibnamefont
  {Mrudul}}, \bibinfo {author} {\bibfnamefont {A.}~\bibnamefont {Pattanayak}},
  \bibinfo {author} {\bibfnamefont {M.}~\bibnamefont {Ivanov}},\ and\ \bibinfo
  {author} {\bibfnamefont {G.}~\bibnamefont {Dixit}},\ }\bibfield  {title}
  {\bibinfo {title} {Direct numerical observation of real-space recollision in
  high-order harmonic generation from solids},\ }\href@noop {} {\bibfield
  {journal} {\bibinfo  {journal} {Physical Review A}\ }\textbf {\bibinfo
  {volume} {100}},\ \bibinfo {pages} {043420} (\bibinfo {year}
  {2019})}\BibitemShut {NoStop}%
\bibitem [{\citenamefont {Imai}\ \emph {et~al.}(2020)\citenamefont {Imai},
  \citenamefont {Ono},\ and\ \citenamefont {Ishihara}}]{imai2020high}%
  \BibitemOpen
  \bibfield  {author} {\bibinfo {author} {\bibfnamefont {S.}~\bibnamefont
  {Imai}}, \bibinfo {author} {\bibfnamefont {A.}~\bibnamefont {Ono}},\ and\
  \bibinfo {author} {\bibfnamefont {S.}~\bibnamefont {Ishihara}},\ }\bibfield
  {title} {\bibinfo {title} {High harmonic generation in a correlated electron
  system},\ }\href@noop {} {\bibfield  {journal} {\bibinfo  {journal} {Physical
  Review Letters}\ }\textbf {\bibinfo {volume} {124}},\ \bibinfo {pages}
  {157404} (\bibinfo {year} {2020})}\BibitemShut {NoStop}%
\bibitem [{\citenamefont {Borsch}\ \emph {et~al.}(2020)\citenamefont {Borsch},
  \citenamefont {Schmid}, \citenamefont {Weigl}, \citenamefont {Schlauderer},
  \citenamefont {Hofmann}, \citenamefont {Lange}, \citenamefont {Steiner},
  \citenamefont {Koch}, \citenamefont {Huber},\ and\ \citenamefont
  {Kira}}]{borsch2020super}%
  \BibitemOpen
  \bibfield  {author} {\bibinfo {author} {\bibfnamefont {M.}~\bibnamefont
  {Borsch}}, \bibinfo {author} {\bibfnamefont {C.~P.}\ \bibnamefont {Schmid}},
  \bibinfo {author} {\bibfnamefont {L.}~\bibnamefont {Weigl}}, \bibinfo
  {author} {\bibfnamefont {S.}~\bibnamefont {Schlauderer}}, \bibinfo {author}
  {\bibfnamefont {N.}~\bibnamefont {Hofmann}}, \bibinfo {author} {\bibfnamefont
  {C.}~\bibnamefont {Lange}}, \bibinfo {author} {\bibfnamefont {J.~T.}\
  \bibnamefont {Steiner}}, \bibinfo {author} {\bibfnamefont {S.~W.}\
  \bibnamefont {Koch}}, \bibinfo {author} {\bibfnamefont {R.}~\bibnamefont
  {Huber}},\ and\ \bibinfo {author} {\bibfnamefont {M.}~\bibnamefont {Kira}},\
  }\bibfield  {title} {\bibinfo {title} {Super-resolution lightwave tomography
  of electronic bands in quantum materials},\ }\href@noop {} {\bibfield
  {journal} {\bibinfo  {journal} {Science}\ }\textbf {\bibinfo {volume}
  {370}},\ \bibinfo {pages} {1204} (\bibinfo {year} {2020})}\BibitemShut
  {NoStop}%
\bibitem [{\citenamefont {Rana}\ \emph {et~al.}(2022)\citenamefont {Rana},
  \citenamefont {Mrudul}, \citenamefont {Kartashov}, \citenamefont {Ivanov},\
  and\ \citenamefont {Dixit}}]{rana2022high}%
  \BibitemOpen
  \bibfield  {author} {\bibinfo {author} {\bibfnamefont {N.}~\bibnamefont
  {Rana}}, \bibinfo {author} {\bibfnamefont {M.}~\bibnamefont {Mrudul}},
  \bibinfo {author} {\bibfnamefont {D.}~\bibnamefont {Kartashov}}, \bibinfo
  {author} {\bibfnamefont {M.}~\bibnamefont {Ivanov}},\ and\ \bibinfo {author}
  {\bibfnamefont {G.}~\bibnamefont {Dixit}},\ }\bibfield  {title} {\bibinfo
  {title} {High-harmonic spectroscopy of coherent lattice dynamics in
  graphene},\ }\href@noop {} {\bibfield  {journal} {\bibinfo  {journal}
  {Physical Review B}\ }\textbf {\bibinfo {volume} {106}},\ \bibinfo {pages}
  {064303} (\bibinfo {year} {2022})}\BibitemShut {NoStop}%
\bibitem [{\citenamefont {Ghimire}\ \emph {et~al.}(2011)\citenamefont
  {Ghimire}, \citenamefont {DiChiara}, \citenamefont {Sistrunk}, \citenamefont
  {Agostini}, \citenamefont {DiMauro},\ and\ \citenamefont
  {Reis}}]{ghimire2011observation}%
  \BibitemOpen
  \bibfield  {author} {\bibinfo {author} {\bibfnamefont {S.}~\bibnamefont
  {Ghimire}}, \bibinfo {author} {\bibfnamefont {A.~D.}\ \bibnamefont
  {DiChiara}}, \bibinfo {author} {\bibfnamefont {E.}~\bibnamefont {Sistrunk}},
  \bibinfo {author} {\bibfnamefont {P.}~\bibnamefont {Agostini}}, \bibinfo
  {author} {\bibfnamefont {L.~F.}\ \bibnamefont {DiMauro}},\ and\ \bibinfo
  {author} {\bibfnamefont {D.~A.}\ \bibnamefont {Reis}},\ }\bibfield  {title}
  {\bibinfo {title} {Observation of high-order harmonic generation in a bulk
  crystal},\ }\href@noop {} {\bibfield  {journal} {\bibinfo  {journal} {Nature
  Physics}\ }\textbf {\bibinfo {volume} {7}},\ \bibinfo {pages} {138} (\bibinfo
  {year} {2011})}\BibitemShut {NoStop}%
\bibitem [{\citenamefont {Tancogne-Dejean}\ \emph {et~al.}(2017)\citenamefont
  {Tancogne-Dejean}, \citenamefont {M{\"u}cke}, \citenamefont {K{\"a}rtner},\
  and\ \citenamefont {Rubio}}]{tancogne2017ellipticity}%
  \BibitemOpen
  \bibfield  {author} {\bibinfo {author} {\bibfnamefont {N.}~\bibnamefont
  {Tancogne-Dejean}}, \bibinfo {author} {\bibfnamefont {O.~D.}\ \bibnamefont
  {M{\"u}cke}}, \bibinfo {author} {\bibfnamefont {F.~X.}\ \bibnamefont
  {K{\"a}rtner}},\ and\ \bibinfo {author} {\bibfnamefont {A.}~\bibnamefont
  {Rubio}},\ }\bibfield  {title} {\bibinfo {title} {Ellipticity dependence of
  high-harmonic generation in solids originating from coupled intraband and
  interband dynamics},\ }\href@noop {} {\bibfield  {journal} {\bibinfo
  {journal} {Nature Communications}\ }\textbf {\bibinfo {volume} {8}},\
  \bibinfo {pages} {1} (\bibinfo {year} {2017})}\BibitemShut {NoStop}%
\bibitem [{\citenamefont {Klemke}\ \emph {et~al.}(2019)\citenamefont {Klemke},
  \citenamefont {Tancogne-Dejean}, \citenamefont {Rossi}, \citenamefont {Yang},
  \citenamefont {Scheiba}, \citenamefont {Mainz}, \citenamefont {Di~Sciacca},
  \citenamefont {Rubio}, \citenamefont {K{\"a}rtner},\ and\ \citenamefont
  {M{\"u}cke}}]{klemke2019polarization}%
  \BibitemOpen
  \bibfield  {author} {\bibinfo {author} {\bibfnamefont {N.}~\bibnamefont
  {Klemke}}, \bibinfo {author} {\bibfnamefont {N.}~\bibnamefont
  {Tancogne-Dejean}}, \bibinfo {author} {\bibfnamefont {G.~M.}\ \bibnamefont
  {Rossi}}, \bibinfo {author} {\bibfnamefont {Y.}~\bibnamefont {Yang}},
  \bibinfo {author} {\bibfnamefont {F.}~\bibnamefont {Scheiba}}, \bibinfo
  {author} {\bibfnamefont {R.}~\bibnamefont {Mainz}}, \bibinfo {author}
  {\bibfnamefont {G.}~\bibnamefont {Di~Sciacca}}, \bibinfo {author}
  {\bibfnamefont {A.}~\bibnamefont {Rubio}}, \bibinfo {author} {\bibfnamefont
  {F.}~\bibnamefont {K{\"a}rtner}},\ and\ \bibinfo {author} {\bibfnamefont
  {O.}~\bibnamefont {M{\"u}cke}},\ }\bibfield  {title} {\bibinfo {title}
  {Polarization-state-resolved high-harmonic spectroscopy of solids},\
  }\href@noop {} {\bibfield  {journal} {\bibinfo  {journal} {Nature
  Communications}\ }\textbf {\bibinfo {volume} {10}},\ \bibinfo {pages} {1}
  (\bibinfo {year} {2019})}\BibitemShut {NoStop}%
\bibitem [{\citenamefont {Saito}\ \emph {et~al.}(2017)\citenamefont {Saito},
  \citenamefont {Xia}, \citenamefont {Lu}, \citenamefont {Kanai}, \citenamefont
  {Itatani},\ and\ \citenamefont {Ishii}}]{saito2017observation}%
  \BibitemOpen
  \bibfield  {author} {\bibinfo {author} {\bibfnamefont {N.}~\bibnamefont
  {Saito}}, \bibinfo {author} {\bibfnamefont {P.}~\bibnamefont {Xia}}, \bibinfo
  {author} {\bibfnamefont {F.}~\bibnamefont {Lu}}, \bibinfo {author}
  {\bibfnamefont {T.}~\bibnamefont {Kanai}}, \bibinfo {author} {\bibfnamefont
  {J.}~\bibnamefont {Itatani}},\ and\ \bibinfo {author} {\bibfnamefont
  {N.}~\bibnamefont {Ishii}},\ }\bibfield  {title} {\bibinfo {title}
  {Observation of selection rules for circularly polarized fields in
  high-harmonic generation from a crystalline solid},\ }\href@noop {}
  {\bibfield  {journal} {\bibinfo  {journal} {Optica}\ }\textbf {\bibinfo
  {volume} {4}},\ \bibinfo {pages} {1333} (\bibinfo {year} {2017})}\BibitemShut
  {NoStop}%
\bibitem [{\citenamefont {Heinrich}\ \emph {et~al.}(2021)\citenamefont
  {Heinrich}, \citenamefont {Taucer}, \citenamefont {Kfir}, \citenamefont
  {Corkum}, \citenamefont {Staudte}, \citenamefont {Ropers},\ and\
  \citenamefont {Sivis}}]{heinrich2021chiral}%
  \BibitemOpen
  \bibfield  {author} {\bibinfo {author} {\bibfnamefont {T.}~\bibnamefont
  {Heinrich}}, \bibinfo {author} {\bibfnamefont {M.}~\bibnamefont {Taucer}},
  \bibinfo {author} {\bibfnamefont {O.}~\bibnamefont {Kfir}}, \bibinfo {author}
  {\bibfnamefont {P.}~\bibnamefont {Corkum}}, \bibinfo {author} {\bibfnamefont
  {A.}~\bibnamefont {Staudte}}, \bibinfo {author} {\bibfnamefont
  {C.}~\bibnamefont {Ropers}},\ and\ \bibinfo {author} {\bibfnamefont
  {M.}~\bibnamefont {Sivis}},\ }\bibfield  {title} {\bibinfo {title} {Chiral
  high-harmonic generation and spectroscopy on solid surfaces using
  polarization-tailored strong fields},\ }\href@noop {} {\bibfield  {journal}
  {\bibinfo  {journal} {Nature Communications}\ }\textbf {\bibinfo {volume}
  {12}},\ \bibinfo {pages} {1} (\bibinfo {year} {2021})}\BibitemShut {NoStop}%
\bibitem [{\citenamefont {He}\ \emph {et~al.}(2022)\citenamefont {He},
  \citenamefont {Guo}, \citenamefont {Gao},\ and\ \citenamefont
  {Liu}}]{he2022dynamical}%
  \BibitemOpen
  \bibfield  {author} {\bibinfo {author} {\bibfnamefont {Y.-L.}\ \bibnamefont
  {He}}, \bibinfo {author} {\bibfnamefont {J.}~\bibnamefont {Guo}}, \bibinfo
  {author} {\bibfnamefont {F.-Y.}\ \bibnamefont {Gao}},\ and\ \bibinfo {author}
  {\bibfnamefont {X.-S.}\ \bibnamefont {Liu}},\ }\bibfield  {title} {\bibinfo
  {title} {Dynamical symmetry and valley-selective circularly polarized
  high-harmonic generation in monolayer molybdenum disulfide},\ }\href@noop {}
  {\bibfield  {journal} {\bibinfo  {journal} {Physical Review B}\ }\textbf
  {\bibinfo {volume} {105}},\ \bibinfo {pages} {024305} (\bibinfo {year}
  {2022})}\BibitemShut {NoStop}%
\bibitem [{\citenamefont {Mrudul}\ \emph {et~al.}(2021)\citenamefont {Mrudul},
  \citenamefont {Jim{\'e}nez-Gal{\'a}n}, \citenamefont {Ivanov},\ and\
  \citenamefont {Dixit}}]{mrudul2021light}%
  \BibitemOpen
  \bibfield  {author} {\bibinfo {author} {\bibfnamefont {M.}~\bibnamefont
  {Mrudul}}, \bibinfo {author} {\bibfnamefont {{\'A}.}~\bibnamefont
  {Jim{\'e}nez-Gal{\'a}n}}, \bibinfo {author} {\bibfnamefont {M.}~\bibnamefont
  {Ivanov}},\ and\ \bibinfo {author} {\bibfnamefont {G.}~\bibnamefont
  {Dixit}},\ }\bibfield  {title} {\bibinfo {title} {Light-induced valleytronics
  in pristine graphene},\ }\href@noop {} {\bibfield  {journal} {\bibinfo
  {journal} {Optica}\ }\textbf {\bibinfo {volume} {8}},\ \bibinfo {pages} {422}
  (\bibinfo {year} {2021})}\BibitemShut {NoStop}%
\bibitem [{\citenamefont {Mrudul}\ and\ \citenamefont
  {Dixit}(2021{\natexlab{a}})}]{mrudul2021controlling}%
  \BibitemOpen
  \bibfield  {author} {\bibinfo {author} {\bibfnamefont {M.~S.}\ \bibnamefont
  {Mrudul}}\ and\ \bibinfo {author} {\bibfnamefont {G.}~\bibnamefont {Dixit}},\
  }\bibfield  {title} {\bibinfo {title} {Controlling valley-polarisation in
  graphene via tailored light pulses},\ }\href@noop {} {\bibfield  {journal}
  {\bibinfo  {journal} {Journal of Physics B}\ }\textbf {\bibinfo {volume}
  {54}},\ \bibinfo {pages} {224001} (\bibinfo {year}
  {2021}{\natexlab{a}})}\BibitemShut {NoStop}%
\bibitem [{\citenamefont {Hafez}\ \emph {et~al.}(2018)\citenamefont {Hafez},
  \citenamefont {Kovalev}, \citenamefont {Deinert}, \citenamefont {Mics},
  \citenamefont {Green}, \citenamefont {Awari}, \citenamefont {Chen},
  \citenamefont {Germanskiy}, \citenamefont {Lehnert}, \citenamefont {Teichert}
  \emph {et~al.}}]{hafez2018extremely}%
  \BibitemOpen
  \bibfield  {author} {\bibinfo {author} {\bibfnamefont {H.~A.}\ \bibnamefont
  {Hafez}}, \bibinfo {author} {\bibfnamefont {S.}~\bibnamefont {Kovalev}},
  \bibinfo {author} {\bibfnamefont {J.~C.}\ \bibnamefont {Deinert}}, \bibinfo
  {author} {\bibfnamefont {Z.}~\bibnamefont {Mics}}, \bibinfo {author}
  {\bibfnamefont {B.}~\bibnamefont {Green}}, \bibinfo {author} {\bibfnamefont
  {N.}~\bibnamefont {Awari}}, \bibinfo {author} {\bibfnamefont
  {M.}~\bibnamefont {Chen}}, \bibinfo {author} {\bibfnamefont {S.}~\bibnamefont
  {Germanskiy}}, \bibinfo {author} {\bibfnamefont {U.}~\bibnamefont {Lehnert}},
  \bibinfo {author} {\bibfnamefont {J.}~\bibnamefont {Teichert}}, \emph
  {et~al.},\ }\bibfield  {title} {\bibinfo {title} {Extremely efficient
  terahertz high-harmonic generation in graphene by hot dirac fermions},\
  }\href@noop {} {\bibfield  {journal} {\bibinfo  {journal} {Nature}\ }\textbf
  {\bibinfo {volume} {561}},\ \bibinfo {pages} {507} (\bibinfo {year}
  {2018})}\BibitemShut {NoStop}%
\bibitem [{\citenamefont {Avetissian}\ and\ \citenamefont
  {Mkrtchian}(2018)}]{avetissian2018impact}%
  \BibitemOpen
  \bibfield  {author} {\bibinfo {author} {\bibfnamefont {H.~K.}\ \bibnamefont
  {Avetissian}}\ and\ \bibinfo {author} {\bibfnamefont {G.~F.}\ \bibnamefont
  {Mkrtchian}},\ }\bibfield  {title} {\bibinfo {title} {Impact of
  electron-electron coulomb interaction on the high harmonic generation process
  in graphene},\ }\href@noop {} {\bibfield  {journal} {\bibinfo  {journal}
  {Physical Review B}\ }\textbf {\bibinfo {volume} {97}},\ \bibinfo {pages}
  {115454} (\bibinfo {year} {2018})}\BibitemShut {NoStop}%
\bibitem [{\citenamefont {Chizhova}\ \emph {et~al.}(2017)\citenamefont
  {Chizhova}, \citenamefont {Libisch},\ and\ \citenamefont
  {Burgd{\"o}rfer}}]{chizhova2017high}%
  \BibitemOpen
  \bibfield  {author} {\bibinfo {author} {\bibfnamefont {L.~A.}\ \bibnamefont
  {Chizhova}}, \bibinfo {author} {\bibfnamefont {F.}~\bibnamefont {Libisch}},\
  and\ \bibinfo {author} {\bibfnamefont {J.}~\bibnamefont {Burgd{\"o}rfer}},\
  }\bibfield  {title} {\bibinfo {title} {High-harmonic generation in graphene:
  Interband response and the harmonic cutoff},\ }\href@noop {} {\bibfield
  {journal} {\bibinfo  {journal} {Physical Review B}\ }\textbf {\bibinfo
  {volume} {95}},\ \bibinfo {pages} {085436} (\bibinfo {year}
  {2017})}\BibitemShut {NoStop}%
\bibitem [{\citenamefont {Al-Naib}\ \emph {et~al.}(2014)\citenamefont
  {Al-Naib}, \citenamefont {Sipe},\ and\ \citenamefont {Dignam}}]{al2014high}%
  \BibitemOpen
  \bibfield  {author} {\bibinfo {author} {\bibfnamefont {I.}~\bibnamefont
  {Al-Naib}}, \bibinfo {author} {\bibfnamefont {J.~E.}\ \bibnamefont {Sipe}},\
  and\ \bibinfo {author} {\bibfnamefont {M.~M.}\ \bibnamefont {Dignam}},\
  }\bibfield  {title} {\bibinfo {title} {High harmonic generation in undoped
  graphene: Interplay of inter-and intraband dynamics},\ }\href@noop {}
  {\bibfield  {journal} {\bibinfo  {journal} {Physical Review B}\ }\textbf
  {\bibinfo {volume} {90}},\ \bibinfo {pages} {245423} (\bibinfo {year}
  {2014})}\BibitemShut {NoStop}%
\bibitem [{\citenamefont {Zurr{\'o}n}\ \emph {et~al.}(2018)\citenamefont
  {Zurr{\'o}n}, \citenamefont {Pic{\'o}n},\ and\ \citenamefont
  {Plaja}}]{zurron2018theory}%
  \BibitemOpen
  \bibfield  {author} {\bibinfo {author} {\bibfnamefont {{\'O}.}~\bibnamefont
  {Zurr{\'o}n}}, \bibinfo {author} {\bibfnamefont {A.}~\bibnamefont
  {Pic{\'o}n}},\ and\ \bibinfo {author} {\bibfnamefont {L.}~\bibnamefont
  {Plaja}},\ }\bibfield  {title} {\bibinfo {title} {Theory of high-order
  harmonic generation for gapless graphene},\ }\href@noop {} {\bibfield
  {journal} {\bibinfo  {journal} {New Journal of Physics}\ }\textbf {\bibinfo
  {volume} {20}},\ \bibinfo {pages} {053033} (\bibinfo {year}
  {2018})}\BibitemShut {NoStop}%
\bibitem [{\citenamefont {Zurr{\'o}n-Cifuentes}\ \emph
  {et~al.}(2019)\citenamefont {Zurr{\'o}n-Cifuentes}, \citenamefont
  {Boyero-Garc{\'\i}a}, \citenamefont {Hern{\'a}ndez-Garc{\'\i}a},
  \citenamefont {Pic{\'o}n},\ and\ \citenamefont {Plaja}}]{zurron2019optical}%
  \BibitemOpen
  \bibfield  {author} {\bibinfo {author} {\bibfnamefont {{\'O}.}~\bibnamefont
  {Zurr{\'o}n-Cifuentes}}, \bibinfo {author} {\bibfnamefont {R.}~\bibnamefont
  {Boyero-Garc{\'\i}a}}, \bibinfo {author} {\bibfnamefont {C.}~\bibnamefont
  {Hern{\'a}ndez-Garc{\'\i}a}}, \bibinfo {author} {\bibfnamefont
  {A.}~\bibnamefont {Pic{\'o}n}},\ and\ \bibinfo {author} {\bibfnamefont
  {L.}~\bibnamefont {Plaja}},\ }\bibfield  {title} {\bibinfo {title} {Optical
  anisotropy of non-perturbative high-order harmonic generation in gapless
  graphene},\ }\href@noop {} {\bibfield  {journal} {\bibinfo  {journal} {Optics
  Express}\ }\textbf {\bibinfo {volume} {27}},\ \bibinfo {pages} {7776}
  (\bibinfo {year} {2019})}\BibitemShut {NoStop}%
\bibitem [{\citenamefont {Taucer}\ \emph {et~al.}(2017)\citenamefont {Taucer},
  \citenamefont {Hammond}, \citenamefont {Corkum}, \citenamefont {Vampa},
  \citenamefont {Couture}, \citenamefont {Thir{\'e}}, \citenamefont {Schmidt},
  \citenamefont {L{\'e}gar{\'e}}, \citenamefont {Selvi}, \citenamefont
  {Unsuree} \emph {et~al.}}]{taucer2017nonperturbative}%
  \BibitemOpen
  \bibfield  {author} {\bibinfo {author} {\bibfnamefont {M.}~\bibnamefont
  {Taucer}}, \bibinfo {author} {\bibfnamefont {T.~J.}\ \bibnamefont {Hammond}},
  \bibinfo {author} {\bibfnamefont {P.~B.}\ \bibnamefont {Corkum}}, \bibinfo
  {author} {\bibfnamefont {G.}~\bibnamefont {Vampa}}, \bibinfo {author}
  {\bibfnamefont {C.}~\bibnamefont {Couture}}, \bibinfo {author} {\bibfnamefont
  {N.}~\bibnamefont {Thir{\'e}}}, \bibinfo {author} {\bibfnamefont {B.~E.}\
  \bibnamefont {Schmidt}}, \bibinfo {author} {\bibfnamefont {F.}~\bibnamefont
  {L{\'e}gar{\'e}}}, \bibinfo {author} {\bibfnamefont {H.}~\bibnamefont
  {Selvi}}, \bibinfo {author} {\bibfnamefont {N.}~\bibnamefont {Unsuree}},
  \emph {et~al.},\ }\bibfield  {title} {\bibinfo {title} {Nonperturbative
  harmonic generation in graphene from intense midinfrared pulsed light},\
  }\href@noop {} {\bibfield  {journal} {\bibinfo  {journal} {Physical Review
  B}\ }\textbf {\bibinfo {volume} {96}},\ \bibinfo {pages} {195420} (\bibinfo
  {year} {2017})}\BibitemShut {NoStop}%
\bibitem [{\citenamefont {Chen}\ and\ \citenamefont
  {Qin}(2019)}]{chen2019circularly}%
  \BibitemOpen
  \bibfield  {author} {\bibinfo {author} {\bibfnamefont {Z.~Y.}\ \bibnamefont
  {Chen}}\ and\ \bibinfo {author} {\bibfnamefont {R.}~\bibnamefont {Qin}},\
  }\bibfield  {title} {\bibinfo {title} {Circularly polarized extreme
  ultraviolet high harmonic generation in graphene},\ }\href@noop {} {\bibfield
   {journal} {\bibinfo  {journal} {Optics Express}\ }\textbf {\bibinfo {volume}
  {27}},\ \bibinfo {pages} {3761} (\bibinfo {year} {2019})}\BibitemShut
  {NoStop}%
\bibitem [{\citenamefont {Sato}\ \emph {et~al.}(2021)\citenamefont {Sato},
  \citenamefont {Hirori}, \citenamefont {Sanari}, \citenamefont {Kanemitsu},\
  and\ \citenamefont {Rubio}}]{sato2021high}%
  \BibitemOpen
  \bibfield  {author} {\bibinfo {author} {\bibfnamefont {S.~A.}\ \bibnamefont
  {Sato}}, \bibinfo {author} {\bibfnamefont {H.}~\bibnamefont {Hirori}},
  \bibinfo {author} {\bibfnamefont {Y.}~\bibnamefont {Sanari}}, \bibinfo
  {author} {\bibfnamefont {Y.}~\bibnamefont {Kanemitsu}},\ and\ \bibinfo
  {author} {\bibfnamefont {A.}~\bibnamefont {Rubio}},\ }\bibfield  {title}
  {\bibinfo {title} {High-order harmonic generation in graphene: Nonlinear
  coupling of intraband and interband transitions},\ }\href@noop {} {\bibfield
  {journal} {\bibinfo  {journal} {Physical Review B}\ }\textbf {\bibinfo
  {volume} {103}},\ \bibinfo {pages} {L041408} (\bibinfo {year}
  {2021})}\BibitemShut {NoStop}%
\bibitem [{\citenamefont {Boyero-Garc{\'\i}a}\ \emph
  {et~al.}(2022)\citenamefont {Boyero-Garc{\'\i}a}, \citenamefont
  {Garc{\'\i}a-Cabrera}, \citenamefont {Zurr{\'o}n-Cifuentes}, \citenamefont
  {Hern{\'a}ndez-Garc{\'\i}a},\ and\ \citenamefont {Plaja}}]{boyero2022non}%
  \BibitemOpen
  \bibfield  {author} {\bibinfo {author} {\bibfnamefont {R.}~\bibnamefont
  {Boyero-Garc{\'\i}a}}, \bibinfo {author} {\bibfnamefont {A.}~\bibnamefont
  {Garc{\'\i}a-Cabrera}}, \bibinfo {author} {\bibfnamefont {O.}~\bibnamefont
  {Zurr{\'o}n-Cifuentes}}, \bibinfo {author} {\bibfnamefont {C.}~\bibnamefont
  {Hern{\'a}ndez-Garc{\'\i}a}},\ and\ \bibinfo {author} {\bibfnamefont
  {L.}~\bibnamefont {Plaja}},\ }\bibfield  {title} {\bibinfo {title}
  {Non-classical high harmonic generation in graphene driven by
  linearly-polarized laser pulses},\ }\href@noop {} {\bibfield  {journal}
  {\bibinfo  {journal} {Optics Express}\ }\textbf {\bibinfo {volume} {30}},\
  \bibinfo {pages} {15546} (\bibinfo {year} {2022})}\BibitemShut {NoStop}%
\bibitem [{\citenamefont {Yoshikawa}\ \emph {et~al.}(2017)\citenamefont
  {Yoshikawa}, \citenamefont {Tamaya},\ and\ \citenamefont
  {Tanaka}}]{yoshikawa2017high}%
  \BibitemOpen
  \bibfield  {author} {\bibinfo {author} {\bibfnamefont {N.}~\bibnamefont
  {Yoshikawa}}, \bibinfo {author} {\bibfnamefont {T.}~\bibnamefont {Tamaya}},\
  and\ \bibinfo {author} {\bibfnamefont {K.}~\bibnamefont {Tanaka}},\
  }\bibfield  {title} {\bibinfo {title} {High-harmonic generation in graphene
  enhanced by elliptically polarized light excitation},\ }\href@noop {}
  {\bibfield  {journal} {\bibinfo  {journal} {Science}\ }\textbf {\bibinfo
  {volume} {356}},\ \bibinfo {pages} {736} (\bibinfo {year}
  {2017})}\BibitemShut {NoStop}%
\bibitem [{\citenamefont {Zhang}\ \emph {et~al.}(2021)\citenamefont {Zhang},
  \citenamefont {Li}, \citenamefont {Li}, \citenamefont {Huang}, \citenamefont
  {Lan},\ and\ \citenamefont {Lu}}]{zhang2021orientation}%
  \BibitemOpen
  \bibfield  {author} {\bibinfo {author} {\bibfnamefont {Y.}~\bibnamefont
  {Zhang}}, \bibinfo {author} {\bibfnamefont {L.}~\bibnamefont {Li}}, \bibinfo
  {author} {\bibfnamefont {J.}~\bibnamefont {Li}}, \bibinfo {author}
  {\bibfnamefont {T.}~\bibnamefont {Huang}}, \bibinfo {author} {\bibfnamefont
  {P.}~\bibnamefont {Lan}},\ and\ \bibinfo {author} {\bibfnamefont
  {P.}~\bibnamefont {Lu}},\ }\bibfield  {title} {\bibinfo {title} {Orientation
  dependence of high-order harmonic generation in graphene},\ }\href@noop {}
  {\bibfield  {journal} {\bibinfo  {journal} {Physical Review A}\ }\textbf
  {\bibinfo {volume} {104}},\ \bibinfo {pages} {033110} (\bibinfo {year}
  {2021})}\BibitemShut {NoStop}%
\bibitem [{\citenamefont {Liu}\ \emph {et~al.}(2018)\citenamefont {Liu},
  \citenamefont {Zheng}, \citenamefont {Zeng},\ and\ \citenamefont
  {Li}}]{liu2018driving}%
  \BibitemOpen
  \bibfield  {author} {\bibinfo {author} {\bibfnamefont {C.}~\bibnamefont
  {Liu}}, \bibinfo {author} {\bibfnamefont {Y.}~\bibnamefont {Zheng}}, \bibinfo
  {author} {\bibfnamefont {Z.}~\bibnamefont {Zeng}},\ and\ \bibinfo {author}
  {\bibfnamefont {R.}~\bibnamefont {Li}},\ }\bibfield  {title} {\bibinfo
  {title} {Driving-laser ellipticity dependence of high-order harmonic
  generation in graphene},\ }\href@noop {} {\bibfield  {journal} {\bibinfo
  {journal} {Physical Review A}\ }\textbf {\bibinfo {volume} {97}},\ \bibinfo
  {pages} {063412} (\bibinfo {year} {2018})}\BibitemShut {NoStop}%
\bibitem [{\citenamefont {Dong}\ \emph {et~al.}(2021)\citenamefont {Dong},
  \citenamefont {Xia},\ and\ \citenamefont {Liu}}]{dong2021ellipticity}%
  \BibitemOpen
  \bibfield  {author} {\bibinfo {author} {\bibfnamefont {F.}~\bibnamefont
  {Dong}}, \bibinfo {author} {\bibfnamefont {Q.}~\bibnamefont {Xia}},\ and\
  \bibinfo {author} {\bibfnamefont {J.}~\bibnamefont {Liu}},\ }\bibfield
  {title} {\bibinfo {title} {Ellipticity of the harmonic emission from graphene
  irradiated by a linearly polarized laser},\ }\href@noop {} {\bibfield
  {journal} {\bibinfo  {journal} {Physical Review A}\ }\textbf {\bibinfo
  {volume} {104}},\ \bibinfo {pages} {033119} (\bibinfo {year}
  {2021})}\BibitemShut {NoStop}%
\bibitem [{\citenamefont {Mrudul}\ and\ \citenamefont
  {Dixit}(2021{\natexlab{b}})}]{mrudul2021high}%
  \BibitemOpen
  \bibfield  {author} {\bibinfo {author} {\bibfnamefont {M.}~\bibnamefont
  {Mrudul}}\ and\ \bibinfo {author} {\bibfnamefont {G.}~\bibnamefont {Dixit}},\
  }\bibfield  {title} {\bibinfo {title} {High-harmonic generation from
  monolayer and bilayer graphene},\ }\href@noop {} {\bibfield  {journal}
  {\bibinfo  {journal} {Physical Review B}\ }\textbf {\bibinfo {volume}
  {103}},\ \bibinfo {pages} {094308} (\bibinfo {year}
  {2021}{\natexlab{b}})}\BibitemShut {NoStop}%
\bibitem [{\citenamefont {Alon}\ \emph {et~al.}(1998)\citenamefont {Alon},
  \citenamefont {Averbukh},\ and\ \citenamefont
  {Moiseyev}}]{alon1998selection}%
  \BibitemOpen
  \bibfield  {author} {\bibinfo {author} {\bibfnamefont {O.~E.}\ \bibnamefont
  {Alon}}, \bibinfo {author} {\bibfnamefont {V.}~\bibnamefont {Averbukh}},\
  and\ \bibinfo {author} {\bibfnamefont {N.}~\bibnamefont {Moiseyev}},\
  }\bibfield  {title} {\bibinfo {title} {Selection rules for the high harmonic
  generation spectra},\ }\href@noop {} {\bibfield  {journal} {\bibinfo
  {journal} {Physical Review Letters}\ }\textbf {\bibinfo {volume} {80}},\
  \bibinfo {pages} {3743} (\bibinfo {year} {1998})}\BibitemShut {NoStop}%
\bibitem [{\citenamefont {Higuchi}\ \emph {et~al.}(2017)\citenamefont
  {Higuchi}, \citenamefont {Heide}, \citenamefont {Ullmann}, \citenamefont
  {Weber},\ and\ \citenamefont {Hommelhoff}}]{higuchi2017light}%
  \BibitemOpen
  \bibfield  {author} {\bibinfo {author} {\bibfnamefont {T.}~\bibnamefont
  {Higuchi}}, \bibinfo {author} {\bibfnamefont {C.}~\bibnamefont {Heide}},
  \bibinfo {author} {\bibfnamefont {K.}~\bibnamefont {Ullmann}}, \bibinfo
  {author} {\bibfnamefont {H.~B.}\ \bibnamefont {Weber}},\ and\ \bibinfo
  {author} {\bibfnamefont {P.}~\bibnamefont {Hommelhoff}},\ }\bibfield  {title}
  {\bibinfo {title} {Light-field-driven currents in graphene},\ }\href@noop {}
  {\bibfield  {journal} {\bibinfo  {journal} {Nature}\ }\textbf {\bibinfo
  {volume} {550}},\ \bibinfo {pages} {224} (\bibinfo {year}
  {2017})}\BibitemShut {NoStop}%
\bibitem [{\citenamefont {Heide}\ \emph {et~al.}(2018)\citenamefont {Heide},
  \citenamefont {Higuchi}, \citenamefont {Weber},\ and\ \citenamefont
  {Hommelhoff}}]{heide2018coherent}%
  \BibitemOpen
  \bibfield  {author} {\bibinfo {author} {\bibfnamefont {C.}~\bibnamefont
  {Heide}}, \bibinfo {author} {\bibfnamefont {T.}~\bibnamefont {Higuchi}},
  \bibinfo {author} {\bibfnamefont {H.~B.}\ \bibnamefont {Weber}},\ and\
  \bibinfo {author} {\bibfnamefont {P.}~\bibnamefont {Hommelhoff}},\ }\bibfield
   {title} {\bibinfo {title} {Coherent electron trajectory control in
  graphene},\ }\href@noop {} {\bibfield  {journal} {\bibinfo  {journal}
  {Physical Review Letters}\ }\textbf {\bibinfo {volume} {121}},\ \bibinfo
  {pages} {207401} (\bibinfo {year} {2018})}\BibitemShut {NoStop}%
\bibitem [{\citenamefont {Heide}\ \emph {et~al.}(2022)\citenamefont {Heide},
  \citenamefont {Kobayashi}, \citenamefont {Johnson}, \citenamefont {Liu},
  \citenamefont {Heinz}, \citenamefont {Reis},\ and\ \citenamefont
  {Ghimire}}]{heide2022probing}%
  \BibitemOpen
  \bibfield  {author} {\bibinfo {author} {\bibfnamefont {C.}~\bibnamefont
  {Heide}}, \bibinfo {author} {\bibfnamefont {Y.}~\bibnamefont {Kobayashi}},
  \bibinfo {author} {\bibfnamefont {A.~C.}\ \bibnamefont {Johnson}}, \bibinfo
  {author} {\bibfnamefont {F.}~\bibnamefont {Liu}}, \bibinfo {author}
  {\bibfnamefont {T.~F.}\ \bibnamefont {Heinz}}, \bibinfo {author}
  {\bibfnamefont {D.~A.}\ \bibnamefont {Reis}},\ and\ \bibinfo {author}
  {\bibfnamefont {S.}~\bibnamefont {Ghimire}},\ }\bibfield  {title} {\bibinfo
  {title} {Probing electron-hole coherence in strongly driven 2d materials
  using high-harmonic generation},\ }\href@noop {} {\bibfield  {journal}
  {\bibinfo  {journal} {Optica}\ }\textbf {\bibinfo {volume} {9}},\ \bibinfo
  {pages} {512} (\bibinfo {year} {2022})}\BibitemShut {NoStop}%
\bibitem [{\citenamefont {Klemke}\ \emph {et~al.}(2020)\citenamefont {Klemke},
  \citenamefont {M{\"u}cke}, \citenamefont {Rubio}, \citenamefont
  {K{\"a}rtner},\ and\ \citenamefont {Tancogne-Dejean}}]{klemke2020role}%
  \BibitemOpen
  \bibfield  {author} {\bibinfo {author} {\bibfnamefont {N.}~\bibnamefont
  {Klemke}}, \bibinfo {author} {\bibfnamefont {O.}~\bibnamefont {M{\"u}cke}},
  \bibinfo {author} {\bibfnamefont {A.}~\bibnamefont {Rubio}}, \bibinfo
  {author} {\bibfnamefont {F.~X.}\ \bibnamefont {K{\"a}rtner}},\ and\ \bibinfo
  {author} {\bibfnamefont {N.}~\bibnamefont {Tancogne-Dejean}},\ }\bibfield
  {title} {\bibinfo {title} {Role of intraband dynamics in the generation of
  circularly polarized high harmonics from solids},\ }\href@noop {} {\bibfield
  {journal} {\bibinfo  {journal} {Physical Review B}\ }\textbf {\bibinfo
  {volume} {102}},\ \bibinfo {pages} {104308} (\bibinfo {year}
  {2020})}\BibitemShut {NoStop}%
\end{thebibliography}%

\end{document}